\documentclass[12pt]{article}
\usepackage{amsmath,amssymb}
\usepackage{bm}

\catcode`\@=11

\ifx\documentclass\undefined
 \def\@sect#1#2#3#4#5#6[#7]#8{\ifnum #2>\c@secnumdepth
     \let\@svsec\@empty\else
     \refstepcounter{#1}\edef\@svsec{\csname prefix#1\endcsname
        \csname the#1\endcsname\hskip 1em}\fi
     \@tempskipa #5\relax
      \ifdim \@tempskipa>\z@
        \begingroup #6\relax
          \@hangfrom{\hskip #3\relax\@svsec}{\interlinepenalty \@M #8\par}%
        \endgroup
       \csname #1mark\endcsname{#7}\addcontentsline
         {toc}{#1}{\ifnum #2>\c@secnumdepth \else
                      \protect\numberline{\csname the#1\endcsname}\fi
                    #7}\else
        \def\@svsechd{#6\hskip #3\relax  %% \relax added 2 May 90
                   \@svsec #8\csname #1mark\endcsname
                      {#7}\addcontentsline
                           {toc}{#1}{\ifnum #2>\c@secnumdepth \else
                             \protect\numberline{\csname the#1\endcsname}\fi
                       #7}}\fi
     \@xsect{#5}}
\else
    \def\@seccntformat#1{\csname prefix#1\endcsname
        \csname the#1\endcsname\quad}
\fi

\@addtoreset{equation}{section}   % Makes \section reset 'equation' counter.
\def\theequation{\arabic{section}.\arabic{equation}}

\def\thebibliography#1{\section*{References\@mkboth
 {REFERENCES}{REFERENCES}}\list
 {\leftbibmark\arabic{enumi}\rightbibmark}{
 \settowidth\labelwidth{\leftbibmark #1\rightbibmark}\leftmargin\labelwidth
 \advance\leftmargin\labelsep
 \usecounter{enumi}}
 \def\newblock{\hskip .11em plus .33em minus -.07em}
 \sloppy\clubpenalty4000\widowpenalty4000
 \sfcode`\.=1000\relax}

\def\@citex[#1]#2{\if@filesw\immediate\write\@auxout{\string\citation{#2}}\fi
  \def\@citea{}\@cite{\@for\@citeb:=#2\do
    {\@citea\def\@citea{,\penalty\@m\ }\@ifundefined
       {b@\@citeb}{{\bf ?}\@warning
       {Citation `\@citeb' on page \thepage \space undefined}}%
\hbox{\csname b@\@citeb\endcsname\citemarkdelim}}}{#1}}

\def\@cite#1#2{\leftcitemark{#1 \if@tempswa , #2\fi}\rightcitemark}
\def\leftcitemark{[}
\def\rightcitemark{]}
\def\citemarkdelim{}
\def\leftbibmark{[}
\def\rightbibmark{]}

\catcode`\@=12

\usepackage{amsmath}
\usepackage{amssymb}

\begin{document}

\begin{titlepage}

%\hbox to \hsize{\YUKAWAmark \hfill YITP-96-28}
%\rightline{KUNS 1406}
%\rightline{August 1996(Revised)}

\vspace{2cm}

\begin{center}\large\bf
Evolution of cosmological perturbations \\
in the universe dominated by resonant scalar fields
\end{center}

%\begin{center}\large\bf
%Evolution of the universe \\ 
%dominated by resonant scalar fields
%\end{center}

\bigskip

%\begin{center}
%Hideo Kodama\footnote{email address: kodama@yukawa.kyoto-u.ac.jp} 
%\end{center}

%\begin{center}\it
%Yukawa Institute for Theoretical Physics, Kyoto University, \\
%Kyoto 606-01, Japan\\
%\end{center}

%\begin{center}
%and
%\end{center}

\begin{center}
Takashi Hamazaki\footnote{email address: yj4t-hmzk@asahi-net.or.jp}
\end{center}

\begin{center}\it
%Department of Physics, Faculty of Science, Tokyo Institute of Technology,\\
%Oh-okayama, Meguro, Tokyo 152-0033, Japan\\
Kamiyugi 3-3-4-606 Hachioji-city\\
Tokyo 192-0373 Japan\\
\end{center}

\bigskip
\bigskip
\begin{center}\bf Abstract\end{center}

Recently a Hamiltonian formulation for the evolution of the 
universe dominated by multiple oscillatory scalar fields was 
developed by the present author and was applied to the investigation 
of the evolution of cosmological perturbations on superhorizon 
scales in the case that scalar fields have incommensurable masses.

In the present paper, the analysis is extended to the case in which 
the masses of scalar fields satisfy resonance conditions 
approximately. In this case, the action-angle variables for the 
system can be classified into fast changing variables and slowly 
changing variables. We show that after an appropriate canonical 
transformation, the part of the Hamiltonian that depends on the fast 
changing angle variables can be made negligibly small, so that the 
dynamics of the system can be effectively determined by a truncated 
Hamiltonian that describes a closed dynamics of the slowly changing 
variables. Utilizing this formulation, we show that the system is 
unstable if this truncated Hamiltonian system has hyperbolic fixed 
point and as a consequence, the Bardeen parameter for a perturbation 
of the sytem grows.

PACS number(s):98.80.Cq

\end{titlepage}

\section{Introduction and Summary}

The inflationary universe model is the most successful model in 
explaining the origin of the present cosmological structures such as 
galaxies and clusters of galaxies. 
In this model, quantum fluctuations of an inflaton field, a scalar 
field driving the inflationary expansion, provide seed perturbations 
which grow and form the present cosmological structures by  
gravitational instability. 
During the slow rolling phase of inflation, these seed perturbations 
are streched beyond the Hubble horizon and their wavelengths stay 
larger than the horizon scale until the perturbations come back 
inside the Hubble horizon during the Friedmann stage after the 
inflation. 
The amplitudes of perturbations at this second horizon crossing, 
which have a direct relevance to the CMB anisotropy observations and 
provide the initial condition for detailed astrophysical models of 
galaxy formation, are determined by the so-called Bardeen parameter. 
Hence, in order to obtain information  on  the inflationary stage of 
the universe from observations of the present universe, we have to 
determine the behavior of the Bardeen parameter of perturbations 
during the superhorizon stage.

If the cosmic matter has a regular equation of state and is 
dominated by a single component, this Bardeen parameter is conserved 
with a good accuracy on superhorizon 
scales\cite{Bardeen.J1980}. 
However, in a realistic model, during the period between the first 
horizon crossing in the inflationary regime and the socond horizon 
crossing in the Friedmann regime, the inflaton field oscillates 
coherently around a local minimum of a potential and their energy is 
gradually transformed into matter and radiation which constitute the 
present universe. 
During this reheating phase, the equation of state becomes singular 
periodically \cite{Kodama.H&Hamazaki1996}, and entropy modes can be 
produced\cite{Hamazaki.T&Kodama1996,adivsentropy}. 
Further, in an inflationary model with a multiple-component inflaton 
field, isocurvature modes can appear\cite{adivsentropy,multiscalar}. 
Therefore, the Bardeen parameter may not be converved even 
approximately in realistic models, and a detailed analysis of its 
behavior is mandatory.

On superhorizon scales, this problem can be reduced to the analysis 
of a spatially homogeneous model for the following reasons. First, 
the evolution of cosmological perturbations on superhorizon scales 
is well described by that in the long wavelength limit with a good 
accuracy\cite{Kodama.H&Hamazaki1996}. 
Further, the long wavelength limit of a solution to the perturbation 
equation can be easily constructed from a homogeneous perturbation 
of the background universe model%
\cite{Taruya.A&Nambu1998,Kodama.H&Hamazaki1998,Sasaki.M&Tanaka1998}.
Hence, we only have to analyse a spatially homogeneous system, which 
is much simpler than the analysis of the exact perturbation equation 
with a finite wave number.
If the background system is exactly solvable, this reduction solves 
the problem completely. However, in the case of multiple scalar 
fields in an expanding universe, it is not the case. Further, the 
system exhibits quite complicated oscillatory behavior. 

There is, however, one powerful method to treat such an oscillatory 
system. It is the Hamiltonian formulation in terms of the 
action-angle variables\cite{Arnold.V.I.}. In fact, in a previous 
paper by the present author\cite{Hamazaki.T2002}, we have developed 
a Hamiltonian formulation for a universe model dominated by multiple 
osillatory scalar fields and have shown that it works well at least 
in the case in which the scalar fields have incommensurable masses. 
In this formulation, we introduce an expansion parameter $\epsilon$ 
that represents the ratio of the cosmic expansion rate to the masses 
of the scalar fields, and decompose the Hamiltonian is into an 
unperturbed part that depends only on the action variables and a 
perturbative part that is of the order $\epsilon$ and bounded by a 
constant multiple of $1/t$, where $t$ is a time parameter of the 
system given by $a^{3/2}$ in terms of the cosmic scale factor $a$. 
Next, we look for a canonical transformation that transforms the 
perturbative part to a quantity of higher order with respect to 
$\epsilon$ and $1/t$. As was shown in Ref. \cite{Hamazaki.T2002}, we 
can construct such a canonical transformation iteratively, and by 
repeated applications of such transformations, we can transform the 
perturbative part to a quantity of the order of an arbitrary power 
of $\epsilon$ and $1/t$. Furthermore, the new Hamiltonian system 
obtained by these transformations becomes solvable if the small 
perturbative part is neglected. By this method, we have proved that 
the Bardeen parameter is conserved with a good accuracy under the 
assumption that the masses of the scalar fields do not satisfy  
resonant relations. 

In the reheating phase, a dynamical instability caused by the 
parametric resonance plays a crucial role in the energy transfer 
from a macroscopic homogeneous mode to finite wavelength 
modes\cite{parametricresonance}. 
This instability can have a significant effect on the evolution of 
cosmological perturbations\cite{rescattering,Zibin}.
In fact, a numerical example showing a non-conservation of the 
Bardeen parameter was presented in \cite{adivsentropy}.
In order to treat this problem, it is necessary to extend our 
Hamiltonian formulation to the resonant case. 
From this point of view, in this paper, we undertake this extension 
and with the help of it, we investigate the dynamical behavior of 
the universe dominated by multiple oscillatory scalar fields whose 
masses satisfy a resonance condition at least approximately.

The present paper is organized as follows. First, in the next 
section, on the basis of the paper\cite{Kodama.H&Hamazaki1998}, we 
explain how to construct a solution to the perturbation equation in 
the long wavelength limit from an exactly spatially homogeneous 
perturbation of the background universe model. Then, in \S3,  we put 
the spatially homogeneous system of multiple scalar fields in an 
expanding universe into the Hamiltonian form and introduce the 
action-angle variables.

In \S4, for the case in which the masses of scalar fields satisfy 
resonance conditions approximately, we decompose the action-angle 
variables to fast changing variables and slowly changing variables, 
and reduce the dynamics of the system to that of slowly changing 
variables by a canonical transformation. Then, with the help of this 
formulation, we estimate the growth rate of perturbations of the 
system in the case in which the time parameter $t$ of the system is 
smaller than $1/\epsilon$. On the basis of this estimate and the 
analysis of some soluble examples, we argue that the Bardeen 
parameter of the system can grow if the system has a hyperbolic 
fixed point. Next, in \S5, we analyze the evolution of a 
perturbation of the system in the time range $t>1/\epsilon$ and show 
that the Bardeen parameter is conserved with a good accuracy in this 
time range. \S6 is devoted to discussions. In order to 
make the presentation clear, the proofs of most of the 
mathematical statements and the calculation of the growth rates
of perturbations in concrete models are given the Appendices.

Throughout the paper, the natural units $c=\hbar=1$ are adopted, and 
$8\pi G$ is denoted as $\kappa^2$. Further, the notation adopted in 
the article\cite{Kodama.H&Sasaki1984} is used for 
perturbation variables,  and their definitions are 
sometimes omitted except for those newly defined in this paper.

\section{Evolution of cosmological perturbations in the 
longwavelength  limit}

As mentioned in the introduction, we can determine the dynamical 
behavior of cosmological perturbations on superhorizon scales by 
studying an exactly spatially homogeneous system. Since the 
perturbative analysis of the latter system can be used to determine 
its Lyapunov exponent, which is an index for a dynamical instability 
including chaos of a Hamiltonian system, we can also analyse the 
dynamical instability and integrability of the inflaton dynamics by 
such a study.

In this section, we summarize the main results of the paper 
\cite{Kodama.H&Hamazaki1998} in the case in which the universe is 
dominated by multi-component scalar fields, and explain how to 
construct a solution to the perturbation equation in the 
long-wavelength limit from an exactly homogeneous perturbation of 
the model. We assume that the universe is spatially flat($K=0$) 
throughout the paper. Hence, the background metric is given by
\begin{equation}
ds^2 = -d\tau ^2 + a(\tau )^2 d\bm{x}^2.
\end{equation}
We consider the universe dominated by multi-component scalar fields 
whose energy-momentum tensor is given by 
\begin{equation}
T^\mu_\nu=\nabla^\mu\phi\cdot\nabla_\nu\phi
-\frac{1}{2}\delta^\mu_\nu
\left(\nabla^\lambda\phi\cdot\nabla_\lambda\phi + 2U\right).
\end{equation}

In the long wavelength limit, the gauge-invariant variable $Y_i$ 
representing the fluctuation of the scalar field $\phi_i$ in the 
flat time slice can be expressed as
\begin{eqnarray}
 Y_i &=& {\mit \chi_i} + {\dot\phi_i \over H}\int d\tau {H^2\over 
 2U}W\left({\dot\phi\over H},{\mit \chi}\right);
\nonumber \\
 W(X_1,X_2)&:=& X_1\cdot \dot X_2 - \dot X_1\cdot X_2,
\label{fieldperturbationformula}
\end{eqnarray}
where ${\mit \chi}_i$ is the combination of the exactly homogeneous 
perturbation of $\phi_i$, $\delta\phi_i$, and the perturbation of 
the cosmic scale factor $a$, $\delta a$,  given by
\begin{equation}
 {\mit \chi}_i = \delta \phi_i - {\dot\phi_i \over H} {\delta a \over a}.
\end{equation}
From the equations of motion, it follows that this quantity always 
satisfies the equation
\begin{equation}
{a^3H^2\over U}W\left({\dot\phi\over H},{\mit \chi}\right)={\rm const}.
\label{k0limitWronskian}
\end{equation}

In general, the general solution to the homogenous dynamical system 
can be expressed in terms of the scale factor $a$ and a set of 
integration constants as
\begin{equation}
 \phi_i = \phi_i (a, C).
\end{equation}
For this expression, ${\mit \chi}_i$ can be simpliy written as
\begin{equation}
 {\mit \chi}_i = {\partial \phi_i (a, C) \over \partial C}.
\label{scalefactorconstant}
\end{equation}

As explained in the introduciton, one of the most important 
quantities describing the evolution behavior of perturbations is the 
Bardeen parameter defined by
\begin{equation}
 \zeta = {\cal R} - {a H \over k} \sigma_g,
\end{equation}
where ${\cal R}$ and $\sigma_g$ are perturbations of the three 
curvature and the shear of each constant time slice, respectively. 
In the present case, the Bardeen parameter $\zeta$ can be written in 
terms of the gauge-invariant variable $Y_i$ as
 \begin{equation}
 \zeta = - H {\dot{\phi}\cdot Y \over (\dot{\phi})^2}.
\end{equation}
Hence, from (\ref {fieldperturbationformula}), 
$\zeta$ is represented in the long-wavelength limit as
\begin{equation}
 \zeta = - H {\dot{\phi}\cdot {\mit \chi} \over (\dot{\phi})^2}
     - \int d\tau {H^2\over 2U}
         W\left({\dot\phi\over H},{\mit \chi}\right).
\end{equation}
From (\ref {k0limitWronskian}), the second term 
on the right-hand side of this equation 
is proportional to $\int d\tau/ a^3$, 
which rapidly approaches a constant as the cosmic time $\tau$ 
increases. Hence, in order to see whether the Bardeen parameter is 
conserved or not in the superhorizon stage, we can concentrate on 
the first term. For this reason, from this point, we assume that 
$\zeta$ is expressed as
\begin{equation}
 \zeta = - H {\dot{\phi}\cdot {\mit \chi} \over (\dot{\phi})^2}.
\end{equation}

\section{Evolution equations of corresponding exactly homogeneous 
universe}

If we use the cosmic scale factor $a$ as the time variable, the 
action for the homogeneous dynamical system introduced in the 
previous section can be put into the following Hamiltonian form:
\begin{eqnarray}
 S &=& \int \sum_i p_{\phi_i} d \phi_i - h_a da,\\
 h_a &=& { 2 \sqrt{3} \over \kappa } \Omega a^2 \Bigl(
         {1 \over 2} {1 \over \Omega^2 a^6} \sum_i p_{\phi_i}^2
         + U(\phi) \Bigr)^{1/2},\\
  U(\phi) &=& {1 \over 2} \sum_i m^2_i \phi_i^2 
+ U_{\rm int}(\phi),
\end{eqnarray}
where $U_{\rm int}$ is assumed to be a sum of monomials in $\phi$ of 
degrees not less than $3$. After changing the time 
variable to
\begin{equation}
t=\left(\frac{a}{a_0}\right)^{3/2},
\end{equation}
this action can be expressed in terms of the non-dimensional 
canonical variables
\begin{equation}
 \Phi = \frac{\phi}{\phi_0},\quad
 P_{\phi} = \frac{p_{\phi}}{ a_0^3 m_0 \phi_0 \Omega},
\end{equation}
and the dimensionless parameters
\begin{equation}
 \mu_i = \frac{m_i}{m_0},\quad
 \epsilon = \frac{\sqrt{3}}{2} \kappa \phi_0,
\end{equation}
as
\begin{eqnarray}
&S^{'} 
 &=  \frac{S}{\Omega a_0^3 m_0 \phi_0^2} \nonumber \\       
&&=  \int \sum_i P_{\phi_i} d \Phi_i 
            - \frac{2}{\epsilon} 
            \left(
    \frac{1}{2} \frac{1}{t^2} \sum_i P_{\phi_i}^2 
    + \frac{1}{2} t^2 \sum_i \mu^2_i \Phi_i^2 
    + t^2 \frac{1}{m_0^2 \phi_0^2 }U_{\rm int}( \phi_0  \Phi)
            \right)^{1/2} dt.
\end{eqnarray}

Now, let us introduce the action-angle variables $(J_i,\theta_i)$ by 
\begin{eqnarray}
 \Phi_i &=& \sqrt{ {2 \over \mu_i} } \sqrt{J_i} {1 \over t} 
 \cos{\theta_i},\\
 P_{\phi_i} &=& - \sqrt{ 2 \mu_i } \sqrt{J_i} t \sin{\theta_i},
\end{eqnarray}
which corresponds to the canonical transformation generated 
by the generating function 
\begin{equation}
 W(\Phi, \theta, t) = - {1 \over 2} 
 \sum_i \mu_i t^2 \Phi_i^2 \tan{\theta_i}.
\end{equation}
Then, the Hamiltonian of the system is transformed to 
\begin{equation}
 H = {2 \over \epsilon} 
   \Bigl(\sum_i \mu_i J_i 
        + {t^2 \over m_0^2 \phi_0^2} U_{\rm int}
   \Bigr)^{1/2}
   - \sum_i {J_i \over t} \sin{2 \theta_i}.
\label{StartingHamiltonian}
\end{equation}
Further, in terms of these new variable, the Bardeen parameter is 
expressed as
\begin{eqnarray}
 \zeta &=& {2 \over 3} {\epsilon \over t} 
       (\mu \cdot J + {t^2 \over m_0^2 \phi_0^2} U_{\rm int}
       )^{1/2} 
       {1 \over \sum_i \mu_i J^i (1 - \cos{2 \theta^i})}
\nonumber \\
   && \sum_i \Bigl(
        {1 \over 2} \delta J^i \sin{2 \theta^i} 
         - J^i (1 - \cos{2 \theta^i}) \delta \theta^i
              \Bigr). 
\end{eqnarray}

\section{ Evolution for $t \le 1/ \epsilon$ }

In this section, we evaluate the growth rate of a perturbation of 
the system defined by the Hamiltonian (\ref {StartingHamiltonian}) 
during the period $1\le t \le 1/\epsilon$, 
assuming that the masses of scalar fields satisfy 
a resonant condition approximately. Before going to general 
arguments, we first explain the basic ideas by two simple models 
consisting of two scalar fields.

The first model is defined by 
\begin{equation}
 U_{\rm int} (\phi) = \lambda \phi_1^2 \phi_2,
\qquad
 2 \mu_1 \approx \mu_2.
\end{equation}
If we change the canonical variables from $(\theta_i,J_i)$ ($i=0,1$) 
to $(q_i,p_i)$ ($i=0,1$) by the linear symplectic transformation 
\begin{eqnarray}
 && \theta_1 = q_0 \quad \quad 
    \theta_2 = 2 q_0 + q_1
\nonumber \\
 && J_1 = p_0 - 2 p_1 \quad \quad
    J_2 = p_1,
\end{eqnarray}
we obtain
\begin{equation}
 \mu_1 J_1 + \mu_2 J_2 =
 \omega_0 p_0 + \omega_1 p_1,
\end{equation}
where
\begin{equation}
 \omega_0 = \mu_1 \quad \quad 
 \omega_1 = - 2 \mu_1 + \mu_2.
\end{equation}
$U_{\rm int}$ can be written
\begin{eqnarray}
 {t^2 \over m_0^2 \phi_0^2} U_{\rm int}
  &=&
      {\eta \over t} J_1 J_2^{1/2}
    \bigl\{
      \cos{(2 \theta_1 - \theta_2)} +
      \cos{(2 \theta_1 + \theta_2)} +
      2 \cos {\theta_2}
    \bigr\}
\nonumber \\
  &=&
      {\eta \over t} (p_0 - 2 p_1) p_1^{1/2}
    \bigl\{
      \cos{q_1} +
      \cos{(4 q_0 + q_1)} +
      2 \cos {(2 q_0 + q_1)}
    \bigr\},
\end{eqnarray}
where
\begin{equation}
 \eta = {\lambda \phi_0 \over m_0^2}
        {1 \over \mu_1 \sqrt{\mu_2} \sqrt{2}}.
\end{equation}
Hence, the contribution of this interaction term to the Hamiltonian 
is of order $1/t$, and when averaged over $q_0$ and $q_1$, it 
becomes of order $1/t^2$. It generally holds on the interaction terms 
of the third degree of $\phi$.
This feature will play an important role in the argument in \S5.

The second model is defined by
\begin{equation}
 U_{\rm int} (\phi) = \lambda \phi_1^2 \phi_2^2, 
\quad \quad 
 \mu_1 \approx \mu_2.
\end{equation}
By the linear symplectic transformation 
\begin{eqnarray}
 && \theta_1 = q_0 \quad \quad 
    \theta_2 = q_0 + q_1
\nonumber \\
 && J_1 = p_0 - p_1 \quad \quad
    J_2 = p_1,
\end{eqnarray}
we obtain
\begin{equation}
 \mu_1 J_1 + \mu_2 J_2 =
 \omega_0 p_0 + \omega_1 p_1,
\end{equation}
where
\begin{equation}
 \omega_0 = \mu_1 \quad \quad 
 \omega_1 = - \mu_1 + \mu_2.
\end{equation}
The interaction term is now written
\begin{eqnarray}
  {t^2 \over m_0^2 \phi_0^2} U_{\rm int}
  &=&
{\eta \over t^2} J_1 J_2
    \bigl\{
      1 +
      {1 \over 2} \cos{(2 \theta_1 - 2 \theta_2)} +
      {1 \over 2} \cos{(2 \theta_1 + 2 \theta_2)} +
      \cos {2 \theta_1} +
      \cos {2 \theta_2}
    \bigr\}
\nonumber \\
  &=&
{\eta \over t^2} (p_0-p_1) p_1
\nonumber \\
 &&
    \bigl\{
      1 +
      {1 \over 2} \cos{2 q_1} +
      {1 \over 2} \cos{(4 q_0 + 2 q_1)} +
      \cos {2 q_0} +
      \cos {(2 q_0 + 2 q_1)}
    \bigr\},
\end{eqnarray}
where
\begin{equation}
 \eta = {\lambda \phi_0^2 \over m_0^2}
        {1 \over \mu_1 \mu_2}.
\end{equation}
Hence, in this model, the contribution of this interaction term to 
the Hamiltonian is of order $1/t^2$, and its dominant part does not 
vanish even after it is averaged with respect to $q_0$ and $q_1$, in 
contrast to the first model.

In both of these models, $\omega_1$ represents the deviation 
from the exact resonance, which is given by $2 \mu_1 = \mu_2$ for 
the first model and by $\mu_1 = \mu_2$ for the second model, 
respectively. Hence, $\omega_1$ is much smaller than $\omega_0$, and 
as a consequence, the motion of $q_0$ is much faster than that of 
$q_1$ in general. For this reason, we call $q_0$ and $q_1$ the fast 
angle variable and the slow angle variable, respectively.

This behavior of the variables $q_i$ suggests that the dynamics of 
the slow variable is well described by a Hamiltonian $\bar H$ 
obtained from $H$ by averaging it  with respect to the fast angle 
variable $q_0$:
\begin{equation}
 \bar{H} = {1 \over 2 \pi} \int^{2 \pi}_0 d q_0 H.
\end{equation}   
For the first model, this averaged Hamiltonian can be expressed as
\begin{equation}
 \bar{H} = {2 \over \epsilon} 
   (\bm{\omega} \cdot \bm{p})^{1/2}
   + {\eta \over \epsilon} {1 \over t}
  {1 \over (\bm{\omega} \cdot \bm{p})^{1/2}} (p_0 - 2 p_1) p^{1/2}_1
  \cos{q_1} + 
  O({\eta^2 \over \epsilon} {1 \over t^2}),
\end{equation}
and for the second model, as
\begin{equation}
 \bar{H} = {2 \over \epsilon} 
   (\bm{\omega} \cdot \bm{p})^{1/2}
   + {\eta \over \epsilon} {1 \over t^2}
  {1 \over (\bm{\omega} \cdot \bm{p})^{1/2}} (p_0 - p_1) p_1
  \{ 1+ {1 \over 2} \cos{2 q_1} \} + 
  O({\eta^2 \over \epsilon} {1 \over t^4}),
\end{equation}
where $\bm{\omega} \cdot \bm{p}=\omega_0 p_0+\omega_1 p_1$. 

This simple procedure, however, does not give a useful 
approximation, since the higher-order terms with respect to $1/t$ 
still contains the large parameter $1/\epsilon$, as is seen from the 
above expressions.

Now, we will show for a generic system consisting of $n$ ($\ge2$) 
oscillatory scalar fields that this difficulty can be 
resolved by taking the average after applying an appropriate 
canonical transformation to the Hamiltonian. For that purpose, we 
transform the original action-angle variables to the fast canonical 
variables $(\bm{q}_0, \bm{p}_0)$ and the slow variables $(\bm{q}_1, 
\bm{p}_1)$, as in the above examples. 

For the time being, suppose that the masses of the system satisfy 
$n_1$ resonance relations of the form $\bm{k} \cdot \bm{\mu}=0$, 
where $\bm{k}$ is 
a vector with irreducible integer coefficients. Let $R=(R_{ij})$ be 
a unimodular integral matrix such that its last $n_1$ rows are given 
by the $n_1$ vectors $\bm{k}$ defining the resonant relations, and 
consider the canonical transformation generated by $W = p_i R_{ij} 
\theta_j$, %
\begin{eqnarray}
 q_j &=& {\partial W \over \partial p_j} 
      = R_{jk} \theta_k,\\
 J_j &=& {\partial W \over \partial \theta_j}
      = p_i R_{ij}.
\end{eqnarray}
Here, note that due to the unimodularity of $R$, we can assume that 
each of the new angle variables $q_i$ also has the period $2 \pi$.

Now, let us decompose these new variables $(\bm{q}, \bm{p})$ 
into two sets of variables as
\begin{eqnarray}
&& (\bm{q}_0)_j= q_j,\quad (\bm{p}_0)_j=p_j
   \quad (j=1,\cdots,n_0),\\
&& (\bm{q}_1)_i = q_{n_0+i},\quad (\bm{p}_1)_i=p_{n_0+i}
   \quad (i=1,\cdots,n_1),
\end{eqnarray}
where $n_0+n_1=n$. Then, the set $(\bm{q}_0,\bm{p}_0)$ becomes the 
fast variable and  the set $(\bm{q}_1,\bm{p}_1)$ becomes the slow 
variable, because $\bm{\mu} \cdot \bm{J}$ can be written as 
\begin{equation}
\bm{\mu} \cdot \bm{J}= \bm{p}_0 \cdot \bm{\omega}_0,
\end{equation}
where
\begin{equation}
(\bm{\omega}_0)_i := R_{ij} \mu _j\quad (i=1,\cdots,n_0).
\end{equation}

In the above, we have imposed exact resonant conditions on the mass 
parameters. However, such strong conditions are rarely satisfied. In 
fact, it is known that for $d > n-1$, the set of points 
in the $\bm{\omega}$ space satisfying the condition
\begin{equation}
 \inf_{\bm{k} \not= \bm{0}, \bm{k} \in \mathbb{Z}^n} 
 |\bm{k}|^d |(\bm{k} \cdot \bm{\omega})| =0
\end{equation}
has measure zero, where 
\begin{equation}
  |\bm{k}| := |k_1|+|k_2|+ \cdot \cdot \cdot +|k_n|.
\end{equation}
Therefore, from this point, we only require that the mass parameters 
satisfy resonance conditions approximately. To be precise, we assume 
that 
\begin{eqnarray}
&& \inf_{\bm{k}_0 \not= \bm{0}, \bm{k} \in \mathbb{Z}^n} 
  |\bm{k}|^d |\bm{k}_0 \cdot \bm{\omega}_0 
            + \bm{k}_1 \cdot \bm{\omega}_1| =C>0,
\label{nonresD0}\\
&& \inf_{\bm{k} \not=\bm{0}, \bm{k} \in \mathbb{Z}^n} 
  |\bm{k}|^d |\bm{k}_0 \cdot \bm{\omega}_0 
            + \bm{k}_1 \cdot \bm{\omega}_1| 
   \le \epsilon C,
\label{resappro}
\end{eqnarray}
where $C$ is a constant of order unity. 
When $\bm{\omega}$ satisfies (\ref{nonresD0}) for some positive
constants $d$, $C$, 
we say that $\bm{\omega}$ is of the class $D_0 (d, C)$.

Next, we show that there exists a canonical transformation such that 
in a system obtained by that transformation, the dynamics of the 
slow variables can be determined independent of the fast angle 
variables with a good accuracy. For that purpose, first note that the 
Hamiltonian of the system can be written in terms of the variables 
$\bm{q}_0,\bm{p}_0,\bm{q}_1$ and $\bm{p}_1$ as
\begin{equation}
 H =
 {2 \over \epsilon} 
   (\bm{\omega}_0\cdot \bm{p}_0 + \bm{\omega}_1\cdot \bm{p}_1)^{1/2} 
  +A(\bm{q}_1, \bm{p}, t) + B (\bm{q}, \bm{p}, t),
\end{equation}
by Taylor expanding $H$ with respect to $1/t$,  
where
\begin{equation}
 (\bm{\omega}_1)_i:=R_{n_0+i\, j}\mu_j\quad(i=1,\cdots,n_0).
\end{equation}
We say that this Hamiltonian is of the type 
$C_m(\sigma,M_1,M_2,\rho)$, if the following conditions are 
satisfied for some positive constants $\sigma, M_1,M_2$, and $\rho$:
\begin{itemize}
\item[(i)] $\bm{\omega} \cdot \bm{p}
           =\bm{\omega}_0 \cdot\bm{p}_0
          + \bm{\omega}_1 \cdot\bm{p}_1$ is bounded as 
\begin{equation}
  |\bm{\omega} \cdot \bm{p}| \ge \sigma.
\end{equation}
\item[(ii)] $tA$ can be extended to an analytic function in the 
domain $D_1(\rho)$ in $\mathbb{C}^{n+n_1+1}$ defined by
\begin{eqnarray}
&& D(\rho) := \left\{ \left(\bm{p}, \bm{q}, \frac{1}{t}\right); 
             {\rm Re}\, \bm{p} \in D + \rho, 
            |{\rm Im}\, \bm{p}| \le \rho,
            |{\rm Im}\, \bm{q}| \le \rho, \right.\notag\\
&&  \qquad\qquad\qquad  \left.   \frac{1}{|t|} \le 1+\rho, 
            \left|{\rm Im}\frac{1}{t}\right| \le \rho 
            \right\},\\
&& D_1(\rho):=D(\rho)|_{\bm{q}_0=\bm{0}},
\end{eqnarray}
where $D+\rho$ denotes the $\rho$-neighbourhood of an interval $D$, 
and in this domain, satisfies the inequality
\begin{equation}
 |t A| \le M_1.
\end{equation}
Further, $tA$ is periodic with respect to  $\bm{q}_1$ and real if  
$(\bm{q}_1, \bm{p}, 1/t)$ are real.
\item[(iii)] $t^m B$ can be extended to an analytic function in the 
domain $D(\rho) \subset \mathbb{C}^{2n+1}$ defined above 
and satisfies the inequality
\begin{equation}
 |t^{m} B| \le \epsilon^{m-1} M_2.
\end{equation}
Further, $t^m B$ is periodic with respect to  $\bm{q}$, real if
$(\bm{q}, \bm{p}, 1/t)$ is real, and satisfies 
\begin{equation}
 {1 \over (2 \pi)^{n_0} } 
 \int_0^{2 \pi} \cdot \cdot \cdot \int_0^{2 \pi}
 d^{n_0} \bm{q}_0 t^m B =0.
\end{equation}
\end{itemize}

Under this notation, the following proposition holds.

\paragraph{Proposition $4.1$}
Let $m$ be some positive interger, and consider the Hamitonian 
$H^{(m)}$ written in terms of fast canonical variables 
$(\bm{q}_0^{(m)},\bm{p}_0^{(m)})$ and slow canonical variables 
$(\bm{q}_1^{(m)},\bm{p}_1^{(m)})$ as
\begin{equation} 
 H^{(m)} =  
 {2 \over \epsilon} 
   (\bm{\omega}_0\cdot\bm{p}_0^{(m)} 
    + \bm{\omega}_1\cdot\bm{p}_1^{(m)})^{1/2} 
   + A_m (\bm{q}_1^{(m)}, \bm{p}^{(m)}, t) 
   + B_m (\bm{q}^{(m)}, \bm{p}^{(m)}, t).            
\end{equation}

Suppose that $\bm{\omega}$ is of the class $D_0 (d, C)$
and that this Hamiltonian is of the type 
$C_m(\sigma_m,M_1^{(m)}, M_2^{(m)},\rho_m)$. Then, for any 
$\delta>0$, there exists $\epsilon_0>0$ such that, for an arbitrary 
$\epsilon$ satisfying
\begin{equation}
 0 < \epsilon < \epsilon_0,
\end{equation}
there exists a function $S_m (\bm{q}, \bm{p}, t)$ satisfying 
the following conditions:
\begin{itemize}
\item[(i)] $S_m$ is periodic with respect to $\bm{q}$ and real if 
$(\bm{q}, \bm{p}, t)$ is real.
\item[(ii)] $t^m S_m$ can be extended to an analytic function in the 
multi-dimensional complex domain $D(\rho_{m+1})$, where 
$\rho_{m+1}=\rho_m-\delta$, and in this domain, satisfies the 
inequality
\begin{equation}
 |t^m S_m| \le \epsilon^m L_1^{(m)}
\end{equation}
for some positive constant $L^{(m)}_1$. 
\item[(iii)] Let $H^{(m+1)}(\bm{q}^{(m+1)},\bm{p}^{(m+1)}, t)$ 
be a Hamitonian obtained from $H^{(m)}$ 
by the canonical transformation generated by 
$S_m(\bm{q}^{(m)},\bm{p}^{(m+1)},t)$:
\begin{eqnarray}
&&
 \bm{p}^{(m)} = \bm{p}^{(m+1)} 
         + {\partial S_m \over \partial \bm{q}^{(m)}},
\\
&&
 \bm{q}^{(m+1)} = \bm{q}^{(m)} 
         + {\partial S_m \over \partial \bm{p}^{(m+1)}},
\\
&&
 H^{(m+1)} = H^{(m)} + {\partial S_m \over \partial t}
\\
&& \quad = {2 \over \epsilon} 
             (\bm{\omega} \cdot \bm{p}^{(m+1)})^{1 / 2} 
             + A_{m+1} (\bm{q}_1^{(m+1)}, \bm{p}^{(m+1)}, t) 
             + B_{m+1} (\bm{q}^{(m+1)}, \bm{p}^{(m+1)}, t).
\end{eqnarray}
Then, $H^{(m+1)}$ is of the type 
$C_{m+1}(\sigma_{m+1},M_1^{(m+1)},M_2^{(m+1)},\rho_{m+1})$ for some 
positive constants $\sigma_{m+1}, M_1^{(m+1)}$, and $M_2^{(m+1)}$, 
and the change of the $A$-term in the Hamiltonians satisfies the 
inequality
\begin{equation}
 |t^{m+1} \{  
      A_{m+1} (\bm{q}_1, \bm{p}, t)-
      A_m     (\bm{q}_1, \bm{p}, t)
          \}| \le \frac{\epsilon^m}{2} M_2^{(m+1)},
\end{equation}
for $(\bm{q}_1, \bm{p}, 1/t)\in D_1(\rho_{m+1})$.
\end{itemize}
({\it For the proof, see the appendix A}).

Since the generating function $S_m$ in this proposition is of the 
order $\epsilon^m/t^m$, we have $B_m \sim \epsilon^{m-1} / t^m$ and 
$B_{m+1} \sim \epsilon^{m} / t^{m+1}$. 
This implies that we can make the 
part of the Hamiltonian that depends on the fast variables 
$\bm{q}_0$ aribitrarily small by taking the original Hamiltonian as 
the starting point ($m=1$) and applying canonical tranformations 
given in the proposition repeatedly. Therefore, we can expect that 
the evolution of the slow variables $(\bm{q}_1,\bm{p}_1)$ can be 
determined with a quite good accuracy from an effective Hamiltonian 
$\bar H^{(m)}(\bm{q}_1^{(m)},\bm{p}^{(m)}, t)$ obtained 
by dicarding $B_m$ term from $H^{(m)}$, 
if we take $m$ sufficiently large. In this 
trancated system, $\bm{p}_0^{(m)}$ becomes constant, and the 
behavior of $\bm{q}_0^{(m)}$ can be obtained by a simple time 
integration of a function that has a definite $t$-dependence when 
$\bm{p}_0^{(m)}$ and a corresponding solution for 
$(\bm{q}_1^{(m)},\bm{p}_1^{(m)})$ are given. 
The evolution of the original 
variables can be calculated from this solution by applying the known 
canonical transformation connecting these two sets of variables.

Next, we show that this expectation is correct by estimating the 
errors produced by the trunction. Let us use the symbol $\Delta$ to 
represent the difference of a quantity for the exact system and the 
corresponding quantity for the truncated system. Further, let us 
denote a quantity for the truncated system by the upper case letter 
of the lower case symbol representing the corresponding quantity for 
the exact system. For example, the 
errors of canonical variables are denoted as
\begin{eqnarray}
 && \Delta \bm{Q} = \bm{q} - \bm{Q},\\
 && \Delta \bm{P} = \bm{p} - \bm{P},
\end{eqnarray}
and the errors of solutions to the perturbation equations for both 
systems are denoted as
\begin{eqnarray}
&& \Delta \delta \bm{Q} = \delta \bm{q} - \delta \bm{Q},\\
&& \Delta \delta \bm{P} = \delta \bm{p} - \delta \bm{P}.
\end{eqnarray}
Further, let us denote the sets of slow variables, 
$(\bm{q}_1,\bm{p}_1)$ and $(\bm{Q}_1,\bm{P}_1)$, by $\bm{z}$ and $\bm{Z}$, 
respectively. Finally, for a function $f(t)$, let us define $\|f\|(t)$ by
\begin{equation}
 \|f\| (t) = \sup_{1 \le s \le t} {|f(s)|}.
\end{equation}
Before evaluating $\Delta \bm{Z}$, $\Delta \delta \bm{Z}$
and $\delta \bm{Z}$, we give the general technique to
evaluate an upper bound on the norm of the solution $\bm{X}$
to the first order differential equation:
\begin{equation}
 \frac{d}{dt} \bm{X} = \Omega \bm{X} + \bm{S},
\end{equation}
where $\bm{X}$, $\bm{S}$ are $N$ column vectors and 
$\Omega$ is $N \times N$ matrix.
For the detail, see the appendix of the paper 
\cite{Hamazaki.T&Kodama1996}.
If we define the norm of the solution $\bm{X}$ by
\begin{equation}
 |\bm{X}|^2 := \bm{X}^{\dagger} \bm{X}, 
\end{equation}
we decompose $\Omega$ into a sum of $\Omega_1$, $\Omega_2$:
$\Omega_1$ from the perturbed part is of order $1/t$ and
$\Omega_2$ from the unperturbed part is of order $\epsilon$
and $\lambda_{mi}$ is the largest eigenvalue of the hermitian
matrix $\Omega_{Hi}:= \Omega^{\dagger}_i + \Omega_i$,
$|\bm{X}|$ is bounded as
\begin{equation}
 \frac{d}{dt} |\bm{X}| \le \frac{\lambda}{2} |\bm{X}|
 + |\bm{S}|,
\end{equation}
where
\begin{equation}
 \lambda := \lambda_{m1} + \lambda_{m2}. 
\end{equation}
If $\lambda$ is an eigenvalue of $\Omega_{Hi}$, 
$- \lambda$ is also an eigenvalue of $\Omega_{Hi}$, 
because $\Omega_i$ can be written in the form of $I S_i$
where $S_i$ is the symmetric matrix and $I$ is defined 
by (\ref{hamiltonrotation}).
So $\lambda_{mi}$ are non-negative.
Therefore the norm of the solution is bounded as
\begin{equation}
 |\bm{X}| \le 
 \exp{ \left[ \int_1 \frac{\lambda}{2} dt   
       \right]}
 \left[   
  |\bm{X}(1)| + \int_1 dt |\bm{S}|
 \right].
\end{equation}
In evaluation of the contribution from the source term,
we use
\begin{equation}
 |\sum_{i=1}^{k} \bm{S}_i| \le 
 \sum_{i=1}^{k} |\bm{S}_i|,
\end{equation}
which is obtained from the Cauchy Schwarz inequality
\begin{equation}
 |\bm{A}^{\dagger} \bm{B}| \le |\bm{A}| |\bm{B}|.
\end{equation}

Since the different definition of the norm  
\begin{equation}
|\bm{X}|_m := \max_{1 \le i \le N} {|(\bm{X})_i|} 
\end{equation}
satisfies 
\begin{equation}
 \frac{1}{\sqrt{N}} |\bm{X}| \le |\bm{X}|_m
 \le |\bm{X}|,
\end{equation}
we can identify $|\bm{X}|$ and $|\bm{X}|_m$ assuming 
that the number of degrees of freedom $N$ is of order 
unity, so we simply omit the subscripts $m$.

Then, the truncation error for the fourth-order system can be 
estimated as follows.
From now on we omit constant coefficients of order unity except
$\Gamma$.

\paragraph{Proposition $4.2A$}
Let $\Gamma$ be an upper bound of the eigenvalues of the 
the hermitian matrix $t \Omega_{H1} / 2$ defined by
\begin{eqnarray}
 \Omega_{H1} &:=& \Omega^{\dagger}_1 + \Omega_1, 
\nonumber \\
 \Omega_1 &:=&
 \left(
 \begin{array}{cc}
  \dfrac{\partial^2 A}{\partial \bm{Q}_1 \partial \bm{P}_1}
 & 
  \dfrac{\partial^2 A}{\partial \bm{P}_1 \partial \bm{P}_1}
 \\
  -\dfrac{\partial^2 A}{\partial \bm{Q}_1 \partial \bm{Q}_1}
 & 
  -\dfrac{\partial^2 A}{\partial \bm{P}_1 \partial \bm{Q}_1}
 \end{array}
 \right),
\label{AAE}
\end{eqnarray}
where $A$ is the $A$-term of the fourth-order Hamiltonian 
$H=H^{(4)}$, and all the elements of $\Omega_1$ are bounded
by $M_1^{(4)} / \delta^2 t$ in the domain 
$D_1 (\rho_4 - \delta)$. 
This quantity $\Gamma$ gives an upper bound on the growth 
rates of the errors $\Delta \bm{Q}_1$, $\Delta \bm{P}_1$, 
$\Delta \delta \bm{Q}_1$, and $\Delta \delta \bm{P}_1$, and
the growth of the perturbations 
$\delta \bm{Q}_1$, $\delta \bm{P}_1$.

Let us define $\beta$ by
\begin{equation}
 \beta = \frac{1}{\Gamma+1},
\end{equation}
Then, for an arbitrary positive $\epsilon$, in the time interval
\begin{equation}
 1 \le t \le \frac{1}{\epsilon^{\beta} },
\end{equation}
the truncation errors of canonical variables are given by
\begin{equation}
 |\Delta \bm{P}_0| \le \epsilon^3, \quad
 |\Delta \bm{Z}| \le \epsilon^2, \quad
 \left| \frac{\Delta \bm{Q}_0}{t} \right| \le \epsilon^2,
\label{AAA}
\end{equation}
and the errors of perturbation variables are
\begin{eqnarray}
 \left\| \frac{\Delta \delta \bm{Q}_0}{t} \right\| 
  &\le& \epsilon^2 |\delta \bm{Q}_0 (1)|
      + \epsilon \|\delta \bm{P}_0 \|
      + \epsilon \|\delta \bm{Z} \|,
\nonumber \\
 \|\Delta \delta \bm{P}_0 \| 
  &\le& \epsilon^3 |\delta \bm{Q}_0 (1) |
      + \epsilon^2 \|\delta \bm{P}_0 \|
      + \epsilon^3 \|\delta \bm{Z} \|,
\nonumber \\
 \|\Delta \delta \bm{Z} \| 
  &\le& \epsilon^2 |\delta \bm{Q}_0 (1)|
      + \epsilon \|\delta \bm{P}_0 \|
      + \epsilon \|\delta \bm{Z} \|,
\label{AAB}
\end{eqnarray}
under  the initial conditions:
\begin{eqnarray}
&& \Delta \bm{P}_0 (1) = \Delta \bm{Z} (1) 
 = \Delta \bm{Q}_0 (1) = 0,\\
&&
 \Delta \delta \bm{Q}_0 (1) = 
 \Delta \delta \bm{P}_0 (1) = 
 \Delta \delta \bm{Z} (1) = 0.
\end{eqnarray}
({\it For the proof see Appendix B.1}.)

This proposition shows that the truncation errors can be made small 
in the fourth-order system ($m=4$). In order to obtain the 
information on the behavior of the original variables, we also have 
to estimate how this truncation error affects the original variables 
through the canonical transformation which connects the fourth-order 
system and the original system. The next proposition gives that 
estimate.

\paragraph{Proposition $4.2B$}
The difference between $(\bm{q}^{(1)}, \bm{p}^{(1)})$ obtained
from $(\bm{q}^{(4)}, \bm{p}^{(4)})$ by the canonical transformation
and $(\bm{Q}^{(1)}, \bm{P}^{(1)})$ obtained 
from $(\bm{Q}^{(4)}, \bm{P}^{(4)})$ by the 
same canonical transformation has the upper bound
\begin{equation}
 |\Delta \bm{P}_0^{(1)}| \le \epsilon^3, \quad
 |\Delta \bm{Z}^{(1)}| \le \epsilon^2, \quad
 \left|{\Delta \bm{Q}_0^{(1)} \over t} \right| 
 \le \epsilon^2,
\label{AAC}
\end{equation}
and the corresponding difference in the perturbation variables is 
estimated as
\begin{eqnarray}
 \left\|\frac{\Delta \delta \bm{Q}_0^{(1)}}{t} \right\| 
  &\le& \epsilon^2 |\delta \bm{Q}_0^{(4)} (1)|
      + \epsilon \|\delta \bm{P}_0^{(4)}\|
      + \epsilon \|\delta \bm{Z}^{(4)} \|,
\nonumber \\
 \|\Delta \delta \bm{P}_0^{(1)} \|
  &\le& \epsilon^2 |\delta \bm{Q}_0^{(4)} (1)|
      + \epsilon \|\delta \bm{P}_0^{(4)}\|
      + \epsilon^2 \|\delta \bm{Z}^{(4)}\|,
\nonumber \\
 \|\Delta \delta \bm{Z}^{(1)}\| 
  &\le& \epsilon^2 |\delta \bm{Q}_0^{(4)} (1)|
      + \epsilon \|\delta \bm{P}_0^{(4)} \|
      + \epsilon \|\delta \bm{Z}^{(4)} \|,
\label{AAD}
\end{eqnarray}
in the same time interval as in the previous proposition.
({\it For the proof see Appendix B.2}.)

These two propositions show that the truncation errors can be made 
small even with respect to the original variables, if we truncate 
the system at the fourth-order. This order is minimal in the sense 
that the trunction at a lower-order system produces errors of order 
unity in the perturbation variables. Conversely, if we go to 
higher-order systems, we can obtain a better approximation. Further, 
we can prolong the time interval in which the approximation is good. 
In fact, we can show that in the $m$-th order system with $m\ge4$, 
the same estimates for the truncation errors as in the above 
propositions hold in the interval
\begin{equation}
 1 \le t \le {1 \over \epsilon^{(m-3)/(\Gamma+1)} }.
\end{equation}
In particular, for $m$ larger than $\Gamma+4$, the approximation is 
good in the interval $1<t<1/\epsilon$.

Finally, let us evaluate the growth of perturbations of a truncated 
higher-order system. The Hamiltonian of such a system can be in 
general expressed as
\begin{equation}
 H = {2 \over \epsilon} 
     (\bm{\omega}_0\cdot \bm{P}_0 + \bm{\omega}_1\cdot\bm{P}_1)^{1/2}
    + A(\bm{Q}_1, \bm{P}, t),
\label{ACA}
\end{equation}
where
\begin{equation}
 |t A| \le M, \quad \quad \bm{\omega}_1=O(\epsilon).
\end{equation}
As mentioned before, the fast action variables $\bm{P}_0$ for this 
system become constants of motion, and the equations of motion for 
the slow variables $(\bm{P}_1,\bm{Q}_1)$ do not contain small 
parameter $\epsilon$ essentially. Therefore, in contrast to 
$\bm{P}_0$, the slow action variables $\bm{P}_1$ are not conserved in 
general, and as a consequence, the Bardeen parameter can 
grow considerably in the time interval $1 \le t \le 1/ \epsilon$.
(For estimate of the growth rates of the Bardeen parameter in 
the concrete examples, see the Appendix E.)
This can be confirmed by the following estimate of the upper 
bound of the perturbation variables 
$\delta \bm{Q}$ and $\delta \bm{P}$.

\paragraph{Proposition $4.3$}
In the time interval $1 \le t \le 1 / \epsilon$, 
the perturbation variables $\delta \bm{Q}$ and $\delta \bm{P}$
are bounded as
\begin{eqnarray}
 |\delta \bm{Z}| &\le& t^\Gamma 
 [|\delta \bm{Z}(1)|+ |\delta \bm{P}_0 (1)| (t-1)],\\
 |\delta \bm{P}_0| &=& |\delta \bm{P}_0 (1)|,\\
 |\delta \bm{Q}_0| &\le& |\delta \bm{Q}_0 (1)|
 + {1 \over \epsilon} |\delta \bm{P}_0 (1)| (t-1)
\nonumber \\
 &&
 + |\delta \bm{Z} (1)| {t^{\Gamma+1}-1 \over \Gamma+1}
 + |\delta \bm{P}_0 (1)| 
 ({t^{\Gamma+2}-1 \over \Gamma+2} 
 -{t^{\Gamma+1}-1 \over \Gamma+1}
 ),
\label{ACB}
\end{eqnarray}
where $\Gamma$ is the quantity defined in the proposition 4.2B. 
In particular, the growth rate of the perturbation variables is not 
exponential and at most a power of $t$.

\paragraph{proof}
The evolution of the slow variables $\delta \bm{Q}_1$ 
and $\delta \bm{P}_1$ is 
determined by the equation
\begin{eqnarray}
 {d \over d t}
 \left( 
\begin{array}{c}
 \delta \bm{Q}_1 \\
 \delta \bm{P}_1 
\end{array}
 \right)
  &=&
  \left( 
\begin{array}{cc}
 \dfrac{\partial^2 A}{\partial \bm{Q}_1 \partial \bm{P}_1} & 
 - \dfrac{\bm{\omega}_1 \cdot \bm{\omega}_1}
         { 2 \epsilon (\bm{\omega} \cdot \bm{P})^{3/2} }
 + \dfrac{\partial^2 A}{\partial \bm{P}_1 \partial \bm{P}_1} \\
 - \dfrac{\partial^2 A}{\partial \bm{Q}_1 \partial \bm{Q}_1}& 
 - \dfrac{\partial^2 A}{\partial \bm{P}_1 \partial \bm{Q}_1}
\end{array}
 \right)
 \left( 
\begin{array}{c}
 \delta \bm{Q}_1 \\
 \delta \bm{P}_1 
\end{array}
 \right)
\nonumber \\
 &&
+ \delta \bm{P}_0
 \left( 
\begin{array}{c}
 - \dfrac{\bm{\omega}_1 \cdot \bm{\omega}_0}  
   { 2 \epsilon (\bm{\omega} \cdot \bm{P})^{3/2} }
 + \dfrac{\partial^2 A}{\partial \bm{P}_0 \partial \bm{P}_1} 
\\
 - \dfrac{\partial^2 A}{\partial \bm{P}_0 \partial \bm{Q}_1} 
\end{array}
 \right).
\end{eqnarray}
By evaluating the coefficients in the right-hand side, we obtain
\begin{equation}
 {d \over d t} |\delta \bm{Z}| \le
 ({\Gamma \over t}+\epsilon) |\delta \bm{Z}|
 + |\delta \bm{P}_0 (1)|, 
\end{equation}
where we have used $\delta \bm{P}_0 = \delta \bm{P}_0 (1)$.
By integrating this inequality, we obtain
\begin{equation}
 |\delta \bm{Z}| \le t^\Gamma \exp[\epsilon (t-1)]
 [|\delta \bm{Z}(1)|+ |\delta \bm{P}_0 (1)| (t-1)].
\end{equation}
Since we are considering the time interval 
$1 \le t \le 1/ \epsilon$, we have
\begin{equation}
 |\delta \bm{Z}| \le t^\Gamma 
 [|\delta \bm{Z}(1)|+ |\delta \bm{P}_0 (1)| (t-1)].
\label{ACC}
\end{equation}

Next, $\delta \bm{Q}_0$ obeys the equation
\begin{eqnarray}
 {d \over d t} \delta \bm{Q}_0
 &=&
 - {1 \over 2}
   {\bm{\omega}_0 \over \epsilon}
   {1 \over  (\bm{\omega} \cdot \bm{P})^{3/2} }
   (\bm{\omega}_0 \cdot \delta \bm{P}_0 
  + \bm{\omega}_1 \cdot \delta \bm{P}_1 )
\nonumber \\
 &&
  +
  (\delta \bm{P}_0 \cdot {\partial \over \partial \bm{P}_0}
  +\delta \bm{Q}_1 \cdot {\partial \over \partial \bm{Q}_1}
  +\delta \bm{P}_1 \cdot {\partial \over \partial \bm{P}_1})
 {\partial A \over \partial \bm{P}_0}.
\end{eqnarray}
Integrating the right-hand side of this equation with respect to $t$ 
yields
\begin{eqnarray}
 |\delta \bm{Q}_0| &\le&
 |\delta \bm{Q}_0 (1)|
 + {1 \over \epsilon} \int_1 dt |\delta \bm{P}_0 (1)|
 + \int_1 dt |\delta \bm{P}_1|
 + \int_1 dt {1 \over t} 
   [|\delta \bm{P}_0 (1)|+ |\delta \bm{Z}|]
\nonumber \\
 &\le& 
 |\delta \bm{Q}_0 (1)| 
 + \frac{1}{\epsilon} \int_1 dt |\delta \bm{P}_0 (1)|
 + \int_1 dt |\delta \bm{Z}|.
\end{eqnarray}
From \eqref{ACC} and this equation, we obtain \eqref{ACB}. $\Box$

Next, we show that the above general estimate of the upper bound on 
the growth of perturbations is rather good, by presenting an example 
in which the upper bound is nearly saturated. Let us consider a 
system in which the Hamiltonian flow in the phase space 
$(\bm{Q}_1,\bm{P}_1)$ has a equilibrium fixed point. For simplicity, 
we assume the exact resonance condition $\bm{\omega}_1 = \bm{0}$.  Then, 
in terms of the phase space variable $\bm{Z}$ defined by
\begin{equation}
 \bm{Z}^i = \bm{Q}_1^i \quad (1 \le i \le n_1),
\quad \quad 
 \bm{Z}^{n_1 +i} = \bm{P}_1^i \quad (1 \le i \le n_1),
\end{equation}
the Hamiltonian equations of motion can be written 
\begin{equation}
 {d \bm{Z} \over d t} = I {\partial H \over \partial \bm{Z}},
\end{equation}
where $I$ is the matrix of degree $2n_1$ expressed in terms of the 
unit matrix $E$ of degree $n_1$ as
\begin{equation}
 I = 
 \begin{pmatrix}
 0 & E \\
 -E & 0
\end{pmatrix}.
\label{hamiltonrotation}
\end{equation}

We consider the Hamiltonian of the form
\begin{equation}
 H = 
 {2 \over \epsilon} (\bm{\omega}_0 \cdot \bm{P}_0)^{1/2}
 + {1 \over t^\gamma} a(\bm{Q}_1, \bm{P}),
\end{equation}
where $\gamma=1$ or $\gamma=2$. 
When the resonant interactions come from interaction
terms of the third order in the scalar fields $\phi$,
$\gamma=1$, and when they come from interaction terms
of the fourth order in $\phi$, $\gamma=2$.
We focus on the dynamical behavior near the fixed point 
$(\bm{P}_0 (1), \bm{Z}(1))$ and expand $a(\bm{P}_0, \bm{Z})$
as
\begin{equation}
 a = a_0 + \bm{b}^T \underline{\bm{P}_0} 
         + {1 \over 2} \underline{\bm{P}_0}^T C 
                       \underline{\bm{P}_0}
       + \underline{\bm{Z}}^T D \underline{\bm{P}_0} 
       + {1 \over 2} 
         \underline{\bm{Z}}^T F \underline{\bm{Z}}, 
\end{equation}
dropping the terms of degree not less than $3$ with respect
to deviations from the equilibrium;
$\underline{\bm{P}_0} = \bm{P}_0 - \bm{P}_0 (1)$,
$\underline{\bm{Z}} = \bm{Z} - \bm{Z} (1)$,
where $T$ implies to take  the transposition of a matrix, $a_0$ is a 
constant, $\bm{b}$ and $\bm{P}_0$ are $n_0$-dimensional vectors, $C$ 
is an $n_0 \times n_0$ symmetric matrix, $D$ is a $2 n_1 \times n_0$ 
matrix, and $F$ is a $2 n_1 \times 2 n_1$ symmetric matrix. 
Since $\bm{Z} (1)$ is the fixed points, $a$ does not contain
$\bm{g}^T \underline{\bm{Z}}$.
Then, 
the evolution equations of the slow variables 
$\underline{\bm{Z}}$ can be written in the matrix form as
\begin{equation}
 {d \underline{\bm{Z}} \over d t} = {1 \over t^\gamma} 
  [
I F \underline{\bm{Z}} + I D \underline{\bm{P}_0}
   ].
\end{equation}
Taking the variation of this equation and using the relations
\begin{equation}
 \delta \bm{P}_0 = \delta \bm{P}_0 (1),
\end{equation}
we obtain
\begin{equation}
 {d \over d t} \delta \bm{Z} = {1 \over t^{\gamma}} 
  [
I F \delta \bm{Z} + I D \delta \bm{P}_0 (1)
   ].
\label{ADA}
\end{equation} 

Here, note that the  $2 n_1 \times 2 n_1$ matrix $X:=IF$ satisfies 
the equation
\begin{equation}
 X^T =I X I.
\end{equation}
From this and $I^2=-1$, it follows that the characteristic polynomial
\begin{equation}
 \Delta (\lambda) = \det(X- \lambda E)
\end{equation}
is an even function of $\lambda$:
\begin{equation}
 \Delta (\lambda) = \Delta (- \lambda).
\label{ADB}
\end{equation}
In addition, since $\Delta(\lambda)$ is real polynomial, we have
\begin{equation}
 \Delta (\lambda) = \Delta (\lambda^{\ast}).
\label{ADC}
\end{equation}
Hence, if $a + b i$, ($a,b$ real) is an eigenvalue of $X$, all of 
$\pm a \pm b i$ are also eigenvalues of $X$. From this, it follows 
that if $X$ has an eigenvalue with a non-vanishing real part, the 
fixed point is unstable. We say that such a fixed point is 
hyperbolic. In constrast, if all eigenvalues of $X$ are pure 
imaginary, the fixed point is stable. We say that such a fixed 
point is elliptic. In the case in which the fixed point is neither 
hyperbolic nor elliptic, i.e., some of the eigenvalues of $X$ vanish 
and the other eigenvalues are pure imaginary, the flow around the 
fixed point is stable in the linear analysis but may become unstable 
if higher-order terms are taken into account.

To be precise, this stability argument applies to this autonomous 
systems. In the present case, due to the existence of the 
time-dependent factor $1/t^\gamma$, the real stability depends on 
the value of $\gamma$. Since the system under consideration is 
linear, we can check it directly by solving the equation. 
First, by diagonalizing the matrix $X$ as
\begin{equation}
 S^{-1}X S = \Lambda, 
\end{equation}
where $\Lambda$ is a diagonal matrix, the general solution for 
$\delta Z$ can be written 
\begin{equation}
 \delta \bm{Z} =  S \exp[\int_1 {d t \over t^\gamma} \Lambda] S^{-1}
 \bm{K} -X^{-1} I D \delta \bm{P}_0 (1),
\end{equation}
where
\begin{equation}
 \bm{K}=\delta \bm{Z} (1) + X^{-1} I D \delta \bm{P}_0 (1).
\end{equation}
Next, the equation for $\delta \bm{Q}_0$ is given by
\begin{equation}
 \frac{d}{d t} \delta \bm{Q}_0 =
 - \frac{1}{2} \frac{\bm{\omega}_0}{\epsilon}
 \frac{\bm{\omega}_0 \cdot \delta \bm{P}_0 (1)}
 {(\bm{\omega}_0 \cdot \bm{P}_0 (1))^{1/2}}
 + \frac{1}{t^\gamma} 
  (C \delta \bm{P}_0 (1) + \delta \bm{Z}^T D).
\end{equation}
Hence,  $\delta \bm{Q}_0$ can be expressed as
\begin{equation}
  \delta \bm{Q}_0 =
  \delta \bm{Q}_0 (1)
 - \frac{\bm{\omega}_0}{2\epsilon}
 \frac{\bm{\omega}_0 \cdot \delta \bm{P}_0 (1)}
 {(\bm{\omega}_0 \cdot \bm{P}_0 (1))^{1/2}}(t - 1) + \bm{R},
\end{equation}
where 
\begin{equation}
 \bm{R} = C \delta \bm{P}_0 (1) \int_1 {d t \over t^\gamma} 
   + \int_1 {d t \over t^\gamma} \delta \bm{Z}^T D.
\end{equation}
When $\gamma=1$, inserting the above expression for $\delta \bm{Z}$ 
into this equation, we obtain 
\begin{equation}
 |\bm{R}| \le {t^{\lambda} -1 \over \lambda} + \ln{t},
\end{equation}
for $\lambda > 0$, and
\begin{equation}
 |\bm{R}| \le \ln{t},
\end{equation}
for $\lambda =0$, where $\lambda$ is the maximum of the real parts 
of the eigenvalues of $X$. Thus, $\delta \bm{Q}_0$ is always 
unstable. In contrast, when $\gamma=2$, $|\bm{R}|$ is bounded 
from above as
\begin{equation}
 |\bm{R}| \le e^{\lambda} +1,
\end{equation}
and the stability of $\delta \bm{Q}_0$ depends on the value of 
$\delta\bm{P}_0$.

Finally, we give an exactly soluble example such that there appear 
both hyperbolic fixed points and elliptic fixed points and the phase 
flow pattern is similar to that of the pendulum in a conservative 
field. It is given by the Hamiltonian
\begin{equation}
 H = \frac{2}{\epsilon} (\omega_0  P_0)^{1/2}
 + \frac{1}{ t^2} 
     \left(\frac{1}{2} a P^2_1 + b \cos{Q_1}\right)
   + \frac{c}{t^2},
\end{equation}
where $a$, $b$, $c$ are positive functions of $P_0$. 

Since there exists a conserved quantity,
\begin{equation}
 C = {1 \over 2} a P^2_1 + b \cos{Q_1} = {\rm const},
\label{ADE}
\end{equation}
in this model, we can determine the phase flow pattern in the 
$(Q_1,P_1)$ plane easily. In particular, we find that the phase 
plain is divided into a region of oscillatory motions and a region 
of rotationary motions by the separatrix defined by
\begin{equation}
 P_1 = \pm 2 \sqrt{b \over a} \sin{Q_1 \over 2}.
\label{ADF}
\end{equation}
Further, we find two types of fixed points. 
One is elliptic points, $P_1 = 0, Q_1 = (2k+1)\pi $, where $k$ is an 
integer, each of which is surrounded by flows corresponding to 
oscillatory motions. The other is hyperbolic fixed points,  $P_1 = 
0, Q_1 = 2 \pi k$, where $k$ is an integer. These fixed points are 
connected by the heteroclinic orbits \eqref{ADF}. This implies that 
near the heteroclinic orbits, the perturbations $\delta Q$ and 
$\delta P$, therefore the Bardeen parameter grows.

\section{ Evolution for $t \ge 1/ \epsilon$ }

We have treated the case in which $\bm{\omega}$ is of the 
class $D_0 (d, C)$ and satisfies the resonance relations
at least approximately (\ref {resappro} ) for 
$t \le 1 / \epsilon$.
When there exists deviation from the exact resonance, 
as the time proceeds, the instability characteristic of
resonance disappears and the evolution of the system is 
reduced to the superposition of oscillations with 
amplitudes and frequencies of different orders.
We demonstrate this fact by investigating the case
in which $\bm{\omega}$ is of the classes $D_0 (d, C)$ and
$D (d, \epsilon C)$ for $t \ge 1/ \epsilon$:
we say that $\bm{\omega}$ is of the class $D (d, C)$ if
\begin{equation}
\inf_{\bm{k} \not=\bm{0}, \bm{k} \in \mathbb{Z}^n} 
  |\bm{k}|^d |\bm{k}_0 \cdot \bm{\omega}_0 
            + \bm{k}_1 \cdot \bm{\omega}_1| 
   = C,
\end{equation}
is satisfied for some positive constants $d$, $C$.

By redefining $t$, $H$ by $\epsilon t$, $H / \epsilon$
respectively, we investigate the evolutionary behavior
for $t \ge 1$ of the system defined by the Hamiltonian
of the type $C_{1,1} (\sigma, M_{11}, M_{12}, M_2, \rho)$:
we say that this Hamiltonian 
\begin{eqnarray}
 H &=&
 {2 \over \epsilon^2} 
   (\bm{\omega} \cdot \bm{p})^{1/2} +
  A (\bm{q}_1, \bm{p}, t) +
  B (\bm{q}, \bm{p}, t),\\
  A (\bm{q}_1, \bm{p}, t) &=&
  D (\bm{p}, t)+ E (\bm{q}_1, \bm{p}, t), 
\end{eqnarray}
is of the type $C_{m,l}(\sigma,M_{11},M_{12},M_2,\rho)$, 
if the following conditions are 
satisfied for some positive constants 
$\sigma, M_{11}, M_{12}, M_2$, and $\rho$:
\begin{itemize}
\item[(i)] $\bm{\omega} \cdot \bm{p}
           =\bm{\omega}_0 \cdot\bm{p}_0
          + \bm{\omega}_1 \cdot\bm{p}_1$ is bounded as 
\begin{equation}
  |\bm{\omega} \cdot \bm{p}| \ge \sigma.
\end{equation}
\item[(ii)] $t^2 D$ can be extended to an analytic function in the 
domain $D_2(\rho):=D(\rho)|_{\bm{q}=\bm{0}}$ 
in $\mathbb{C}^{n+1}$  and satisfies the inequality
\begin{equation}
 |t^2 D| \le \epsilon M_{11}.
\end{equation}
Further, $t^2 D$ is real if $(\bm{p}, 1/t)$ are real.
\item[(iii)] $t^l E$ can be extended to an analytic function 
in the domain $D_1(\rho) \subset \mathbb{C}^{n+n_1+1}$ 
and satisfies the inequality
\begin{equation}
 |t^{l} E| \le \epsilon^{l-1} M_{12}.
\end{equation}
Further, $t^l E$ is periodic with respect to  $\bm{q}_1$, 
real if $(\bm{q}_1, \bm{p}, 1/t)$ is real, and satisfies 
\begin{equation}
 {1 \over (2 \pi)^{n_1} } 
 \int_0^{2 \pi} \cdot \cdot \cdot \int_0^{2 \pi}
 d^{n_1} \bm{q}_1 t^l E =0.
\end{equation}
\item[(iv)] $t^m B$ can be extended to an analytic function in the 
domain $D(\rho) \subset \mathbb{C}^{2n+1}$ 
and satisfies the inequality
\begin{equation}
 |t^{m} B| \le \epsilon^{2 (m-1)} M_2.
\end{equation}
Further, $t^m B$ is periodic with respect to  $\bm{q}$, real if
$(\bm{q}, \bm{p}, 1/t)$ is real, and satisfies 
\begin{equation}
 {1 \over (2 \pi)^{n_0} } 
 \int_0^{2 \pi} \cdot \cdot \cdot \int_0^{2 \pi}
 d^{n_0} \bm{q}_0 t^m B =0.
\end{equation}
\end{itemize}

Since the interaction terms of the third degree of $\phi$
do not contribute $D$, $D$ begins with terms of order
$\epsilon / t^2$.

First we show that we can make $\bm{q}_0$ depenent part
$B$ small by the canonical transformation $S$.
We say that the Hamiltonian is of the type
$C^2_m(\sigma,M_1,M_2,\rho)$ if the conditions with $\epsilon$
replaced with $\epsilon^2$ in the definition of the type
$C_m(\sigma,M_1,M_2,\rho)$ are satisfied.

\paragraph{Proposition $5.1$}
Let $m$ be some positive interger, and consider the Hamitonian 
$H^{(m)}$ written in terms of fast canonical variables 
$(\bm{q}_0^{(m)},\bm{p}_0^{(m)})$ and slow canonical variables 
$(\bm{q}_1^{(m)},\bm{p}_1^{(m)})$ as
\begin{equation} 
 H^{(m)} =  
 {2 \over \epsilon^2} 
   (\bm{\omega}_0\cdot\bm{p}_0^{(m)} 
  + \bm{\omega}_1\cdot\bm{p}_1^{(m)})^{1/2} 
   + A_m (\bm{q}_1^{(m)}, \bm{p}^{(m)}, t) 
   + B_m (\bm{q}^{(m)}, \bm{p}^{(m)}, t).            
\end{equation}

Suppose that $\bm{\omega}$ is of the class $D_0 (d, C)$
and that this Hamiltonian is of the type 
$C^2_m(\sigma_m,M_1^{(m)}, M_2^{(m)},\rho_m)$. Then, for any 
$\delta>0$, there exists $\epsilon_0>0$ such that, for an arbitrary 
$\epsilon$ satisfying
\begin{equation}
 0 < \epsilon < \epsilon_0,
\end{equation}
there exists a function $S_m (\bm{q}, \bm{p}, t)$ satisfying 
the following conditions:
\begin{itemize}
\item[(i)] $S_m$ is periodic with respect to $\bm{q}$ and real if 
$(\bm{q}, \bm{p}, t)$ is real.
\item[(ii)] $t^m S_m$ can be extended to an analytic function in the 
multi-dimensional complex domain $D(\rho_{m+1})$, where 
$\rho_{m+1}=\rho_m-\delta$, and in this domain, satisfies the 
inequality
\begin{equation}
 |t^m S_m| \le \epsilon^{2m} L_1^{(m)}
\end{equation}
for some positive constant $L^{(m)}_1$. 
\item[(iii)] Let $H^{(m+1)}(\bm{q}^{(m+1)},\bm{p}^{(m+1)}, t)$ 
be a Hamitonian obtained from $H^{(m)}$ 
by the canonical transformation generated by 
$S_m(\bm{q}^{(m)},\bm{p}^{(m+1)},t)$:
\begin{eqnarray}
&&
 \bm{p}^{(m)} = \bm{p}^{(m+1)} + {\partial S_m \over \partial \bm{q}^{(m)}},
\\
&&
 \bm{q}^{(m+1)} = \bm{q}^{(m)} + {\partial S_m \over \partial \bm{p}^{(m+1)}},
\\
&&
 H^{(m+1)} = H^{(m)} + {\partial S_m \over \partial t}
\\
&& \quad = {2 \over \epsilon^2} 
          (\bm{\omega} \cdot \bm{p}^{(m+1)})^{1 / 2} 
             + A_{m+1} (\bm{q}_1^{(m+1)}, \bm{p}^{(m+1)}, t) 
             + B_{m+1} (\bm{q}^{(m+1)}, \bm{p}^{(m+1)}, t)
\end{eqnarray}
Then, $H^{(m+1)}$ is of the type 
$C^2_{m+1}(\sigma_{m+1},M_1^{(m+1)},M_2^{(m+1)},\rho_{m+1})$ for some 
positive constants $\sigma_{m+1}, M_1^{(m+1)}$, and $M_2^{(m+1)}$, 
and the change of the $A$-term in the Hamiltonians satisfies the 
inequality
\begin{equation}
 |t^{m+1} \{  
      A_{m+1} (\bm{q}_1, \bm{p}, t)-
      A_m     (\bm{q}_1, \bm{p}, t)
          \}| \le \frac{\epsilon^{2m}}{2} M_2^{(m+1)},
\end{equation}
for $(\bm{q}_1, \bm{p}, 1/t)\in D_1(\rho_{m+1})$.
\end{itemize}
The proof is obtained by replacing $\epsilon$ with $\epsilon^2$
in the proof of the Proposition $4.1$.

We show that the fast action variables $\bm{p}_0$ are perpetually
stable and oscillate around the initial values with
amplitudes of order $\epsilon^2$ and with frequencies of order
$1 / \epsilon^2$.

\paragraph{Proposition $5.2$}
There exists some constant $C_0$ such that
\begin{equation}
 |\bm{p}^{(1)}_0 - \bm{p}^{(1)}_0 (1)| \le \epsilon^2 C_0.
\end{equation}

\paragraph{proof}
We consider the transformed Hamiltonian given by
\begin{equation}
 H^{(2)}= {2 \over \epsilon^2} 
          (\bm{\omega} \cdot \bm{p}^{(2)})^{1/2}
         +A_2 + B_2,
\end{equation}
where 
\begin{eqnarray}
 |t A_2| &\le& M^{(2)}_1,\\
 |t^2 B_2| &\le& \epsilon^2 M^{(2)}_2
\end{eqnarray}
which gives
\begin{equation}
 {d \bm{p}^{(2)}_0 \over d t} = 
 - {\partial B_2 \over \partial \bm{q}^{(2)}_0 }.
\end{equation}
We obtain
\begin{equation}
 |\bm{p}^{(2)}_0 - \bm{p}^{(2)}_0 (1)| 
 \le \int_1 dt {\epsilon^2 \over t^2} 
               {M^{(2)}_2 \over \delta}
 \le \epsilon^2 {M^{(2)}_2 \over \delta}.
\end{equation}
By using the transformation law
\begin{equation}
 |\bm{p}^{(1)}-\bm{p}^{(2)}| 
 \le |{\partial S_1 \over \partial \bm{q}^{(1)}}|
 \le {\epsilon^2 \over t} 
     {L^{(1)}_1 \over \delta}, 
\end{equation}
we obtain
\begin{eqnarray}
&&  |\bm{p}^{(1)}_0 - \bm{p}^{(1)}_0 (1)|
\nonumber \\
&& \le |\bm{p}^{(1)}_0 - \bm{p}^{(2)}_0|
    +  |\bm{p}^{(2)}_0 - \bm{p}^{(2)}_0 (1)|
    +  |\bm{p}^{(2)}_0 (1) - \bm{p}^{(1)}_0 (1)|
\nonumber \\
&& \le \epsilon^2  {L^{(1)}_1 \over \delta}
    +  \epsilon^2  {M^{(2)}_2 \over \delta}
    +  \epsilon^2  {L^{(1)}_1 \over \delta},
\end{eqnarray}
which completes the proof.$\Box$

Next we show that we can make $\bm{q}_1$ dependent part $E$
small by the canonical transformation $T$.
\paragraph{Proposition $5.3$}
Let $m$, $l$ be some positive intergers, and consider the Hamitonian 
$H^{(m,l)}$ written in terms of fast canonical variables 
$(\bm{q}_0^{(m,l)},\bm{p}_0^{(m,l)})$ and slow canonical variables 
$(\bm{q}_1^{(m,l)},\bm{p}_1^{(m,l)})$ as
\begin{equation} 
 H^{(m,l)} =  
 {2 \over \epsilon^2} 
   (\bm{\omega}_0\cdot\bm{p}_0^{(m,l)} 
  + \bm{\omega}_1\cdot\bm{p}_1^{(m,l)})^{1/2} 
   + D_{m,l} (\bm{p}^{(m,l)}, t) 
   + E_{m,l} (\bm{q}_1^{(m,l)}, \bm{p}^{(m,l)}, t) 
   + B_{m,l} (\bm{q}^{(m,l)}, \bm{p}^{(m,l)}, t).            
\end{equation}

Suppose that $\bm{\omega}$ is of the class 
$D (d, \epsilon C)$
and that this Hamiltonian is of the type 
$C_{m,l}(\sigma_{m,l},M_{11}^{(m,l)}, M_{12}^{(m,l)},
M_2^{(m,l)},\rho_{m,l})$. Then, for any 
$\delta>0$, there exists $\epsilon_0>0$ such that, for an arbitrary 
$\epsilon$ satisfying
\begin{equation}
 0 < \epsilon < \epsilon_0,
\end{equation}
there exists a function $T_{m,l} (\bm{q}_1, \bm{p}, t)$ satisfying 
the following conditions:
\begin{itemize}
\item[(i)] $T_{m,l}$ is periodic with respect to $\bm{q}_1$ 
and real if $(\bm{q}_1, \bm{p}, t)$ is real.
\item[(ii)] $t^l T_{m,l}$ can be extended to an analytic 
function in the multi-dimensional complex domain 
$D_1 (\rho_{m,l+1})$, 
where $\rho_{m,l+1}=\rho_{m,l}-\delta$, and in this domain, 
satisfies the inequality
\begin{equation}
 |t^l T_{m,l}| \le \epsilon^l L_2^{(m,l)}
\end{equation}
for some positive constant $L^{(m,l)}_2$. 
\item[(iii)] Let $H^{(m,l+1)}(\bm{q}^{(m,l+1)},\bm{p}^{(m,l+1)}, t)$ 
be a Hamitonian obtained from $H^{(m,l)}$ 
by the canonical transformation generated by 
$T_{m,l}(\bm{q}^{(m,l)}_1,\bm{p}^{(m,l+1)},t)$:
\begin{eqnarray}
&&
 \bm{p}_0^{(m,l)} = \bm{p}_0^{(m,l+1)},
\\
&&
 \bm{p}_1^{(m,l)} = \bm{p}_1^{(m,l+1)} 
    + {\partial T_{m,l} \over \partial \bm{q}_1^{(m,l)}},
\\
&&
 \bm{q}^{(m,l+1)} = \bm{q}^{(m,l)} 
    + {\partial T_{m,l} \over \partial \bm{p}^{(m,l+1)}},
\\
&&
 H^{(m,l+1)} = H^{(m,l)} + {\partial T_{m,l} \over \partial t}
\\
&& = {2 \over \epsilon^2} (\bm{\omega} \cdot \bm{p}^{(m,l+1)})^{1 / 2}
             + D_{m,l+1} (\bm{p}^{(m,l+1)}, t) 
             + E_{m,l+1} (\bm{q}_1^{(m,l+1)}, \bm{p}^{(m,l+1)}, t) 
\nonumber \\
&& \quad + B_{m,l+1} (\bm{q}^{(m,l+1)}, \bm{p}^{(m,l+1)}, t)
\end{eqnarray}
Then, $H^{(m,l+1)}$ is of the type 
$C_{m,l+1}(\sigma_{m,l+1},M_{11}^{(m,l+1)},
M_{12}^{(m,l+1)},M_2^{(m,l+1)},\rho_{m,l+1})$ for some 
positive constants $\sigma_{m,l+1}, M_{11}^{(m,l+1)},
M_{12}^{(m,l+1)}$, and $M_2^{(m,l+1)}$, 
and the change of the $D$-term in the Hamiltonians satisfies the 
inequality
\begin{equation}
 |t^{l+1} \{  
      D_{m,l+1} (\bm{p}, t)-
      D_{m,l}   (\bm{p}, t)
          \}| \le \frac{\epsilon^l}{2} M_{12}^{(m,l+1)},
\end{equation}
for $(\bm{p}, 1/t)\in D_2(\rho_{m,l+1})$.
\end{itemize}
({\it For the proof, see the appendix C}).

We show that the slow action variables $\bm{p}_1$ 
are perpetually stable and oscillate around the initial 
values with amplitudes of order $\epsilon$ and with
frequencies of order $1/ \epsilon$.
Notice that $T$ is independent of $\bm{q}_0$. 

\paragraph{Proposition $5.4$}
There exists some constant $C_1$ such that
\begin{equation}
 |\bm{p}^{(1)}_1 - \bm{p}^{(1)}_1 (1)| \le \epsilon C_1.
\end{equation}

\paragraph{proof}
We consider the transformed Hamiltonian $H^{(m, l)}$
($m=2$, $l=2$) given by
\begin{equation}
 H^{(2, 2)}= {2 \over \epsilon^2} 
          (\bm{\omega} \cdot \bm{p}^{(2, 2)})^{1/2}
         +D_{22}+ E_{22}+ B_{22},
\end{equation}
where 
\begin{eqnarray}
 |t^2 D_{22}| &\le& \epsilon M^{(2,2)}_{11},\\
 |t^2 E_{22}| &\le& \epsilon M^{(2,2)}_{12},\\
 |t^2 B_{22}| &\le& \epsilon^2 M^{(2,2)}_2.
\end{eqnarray}
The evolution equations for the slow action variables
$\bm{p}_1$ are given by
\begin{equation}
 {d \bm{p}^{(2,2)}_1 \over d t} = 
 - {\partial E_{22} \over \partial \bm{q}^{(2,2)}_1 }
 - {\partial B_{22} \over \partial \bm{q}^{(2,2)}_1 }
 \sim {\epsilon \over t^2},
\end{equation}
which yields 
\begin{equation}
 |\bm{p}^{(2,2)}_1 - \bm{p}^{(2,2)}_1 (1)| \le \epsilon C,
\end{equation}
for some positive constant $C$.
\begin{eqnarray}
 &&
 |\bm{p}^{(1,1)}_1 (t)- \bm{p}^{(1,1)}_1 (1)| 
\nonumber \\
 &\le& 
 |\bm{p}^{(1,1)}_1 (t)- \bm{p}^{(2,2)}_1 (t)|
+|\bm{p}^{(2,2)}_1 (t)- \bm{p}^{(2,2)}_1 (1)|
+|\bm{p}^{(2,2)}_1 (1)- \bm{p}^{(1,1)}_1 (1)|
\nonumber \\
 &\le& \epsilon C_1.
\end{eqnarray}
$\Box$

We investigate the dynamical behavior of the 
cosmological perturbations.
We obtain the Hamiltonian $H^{(3,3)}$ 
of the type 
$C_{3,3}(\sigma_{3,3},M_{11}^{(3,3)}, M_{12}^{(3,3)},
M_2^{(3,3)},\rho_{3,3})$
by carrying out the canonical transformations in the order 
$S_1$, $S_2$, $T_{3,1}$, $T_{3,2}$. 

Later we omit the superscript $(3,3)$ and constant 
coefficients of order $1$.

From now on we assume that
\begin{equation}
      |\delta \bm{q}_0 (1)|
 \sim |\delta \bm{p}_0 (1)|
 \sim |\delta \bm{q}_1 (1)|
 \sim |\delta \bm{p}_1 (1)|
 \sim 1,
\end{equation}
since in the linear perturbation, the scale of the 
perturbation variables is arbitrary.
Then we obtain the proposition below.

\paragraph{Proposition $5.5$}
The transformed perturbation variables are estimated as
\begin{eqnarray}
&&
 |\delta \bm{p}_0 - \delta \bm{p}_0 (1)| \le \epsilon^2,\\
&&
 |\delta \bm{p}_1 - \delta \bm{p}_1 (1)| \le \epsilon,\\
&&
 \delta \bm{q}_0 = - {1 \over 2} {\bm{\omega}_0 \over \epsilon^2}
     (t - 1) {1 \over (\bm{\omega} \cdot \bm{p}(1))^{3/2}}
      (\bm{\omega}_0 \cdot \delta \bm{p}_0 (1)
      +\bm{\omega}_1 \cdot \delta \bm{p}_1(1))
      + \bm{R}_0,\\
&&
 \delta \bm{q}_1 = - {1 \over 2} {\bm{\omega}_1 \over \epsilon^2}
     (t - 1) {1 \over (\bm{\omega} \cdot \bm{p}(1))^{3/2}}
      (\bm{\omega}_0 \cdot \delta \bm{p}_0 (1)
      +\bm{\omega}_1 \cdot \delta \bm{p}_1(1))
      + \bm{R}_1,
\end{eqnarray}
where the residual part $\bm{R}_0$, $\bm{R}_1$ is bounded as
\begin{eqnarray}
 |\bm{R}_0| &\le& C_0 (t-1),\\
 |\bm{R}_1| &\le& \epsilon C_1 (t-1),
\end{eqnarray}
for some positive constants $C_0$, $C_1$.

{\it For the proof see the appendix (subsection D.1).}

We have evaluated the canonically transformed
variables $\delta \bm{q}^{(3,3)}$, $\delta \bm{p}^{(3,3)}$.
Then we evaluate the difference between such transformed 
variables and the original variables 
$\delta \bm{q}^{(1,1)}$, $\delta \bm{p}^{(1,1)}$.

\paragraph{Proposition $5.6$}
\begin{eqnarray}
 |\delta \bm{q}^{(1,1)}-\delta \bm{q}^{(3,3)}| &\le& 1,\\
 |\delta \bm{p}^{(1,1)}-\delta \bm{p}^{(3,3)}| &\le& 1.
\end{eqnarray}

{\it For the proof see the appendix (subsection D.2).}

According to propositions $5.5$, $5.6$, for $t \ge 1/ \epsilon$,
the Bardeen parameter $\zeta$ stays constant in a good accuracy.

\section{Discussion}

In this paper we have constructed the method for analyzing
the resonance phenomena of the Hamiltonian system obtained
from the multiple oscillatory scalar fields in the expanding
universe.
We have shown that the perturbations including the Bardeen 
parameter can grow when the resonant interactions between 
the homogeneous modes exist. 
If the truncated Hamiltonian system has hyperbolic fixed points
in the phase space of the slow variables, the perturbations 
grow at the speed of a power of $t$ in the three leg
interaction systems, while the growth of perturbations is bounded
from above in no less than four leg interaction systems.
In the models where $\bm{\omega}_1 = 0$, we have found the 
network constructed by the heteroclinic orbits around which 
perturbations are unstable.
On this network, the orbits become irregular, unperiodic,
complicated and probabilistic.
In order to evaluate the amplitude of the Bardeen parameter 
which is directly related the cosmic structure formation,
we must calculate the measure occupied by the stochastic 
network in the whole phase space of the homogeneous modes.

Recentlty the relaxation in reheating is investigated numerically
and is interpreted in terms of the turbulence phenomena of the 
dynamical system of large number of degrees of freedom
\cite{Felder}.
In order to understand the turbulence phenomena,
our fixed point analysis of the Hamiltonian system may be useful.

\section*{Acknowledgments}

The author would like to thank Professor H. Kodama 
for reading his manuscript carefully, helping him improve
his English and giving him useful comments
The author would also like to thank Professor V.I.Arnold 
for his writing the excellent textbooks and/or reviews 
\cite{Arnold.V.I.}, from which the author learned quite a lot about 
the Hamiltonian dynamical system.

\appendix

%% Equation numbering %%
% koko de Appendix no siki banngou wo (A,1) ni suru

\catcode`\@=11

\@addtoreset{equation}{section}   % Makes \section reset 'equation' counter.
\def\theequation{\Alph{section}.\arabic{equation}}

\section{Appendix; Proof of Proposition $4.1$}

We consider the canonical transformation induced by 
the generating function given by
\begin{equation}
 S_m (\bm{p}^{(m+1)}, \bm{q}^{(m)}, t) =
 \sum_{\bm{k}_0 \ne \bm{0}} S_{\bm{k}} (\bm{p}^{(m+1)}, t) 
 e^{i \bm{k} \cdot \bm{q}^{(m)}},
\end{equation}
where 
$\bm{k} \cdot \bm{q}^{(m)}
= \bm{k}_0 \cdot \bm{q}_0^{(m)}+ \bm{k}_1 \cdot \bm{q}_1^{(m)}$
and the sum is taken over $\bm{k}$ such that $\bm{k}_0 \neq \bm{0}$.
$B_m$ is decomposed as 
\begin{equation}
 B_m (\bm{p}^{(m+1)}, \bm{q}^{(m)}, t) = 
 \sum_{\bm{k}_0 \ne \bm{0}} b_{\bm{k}} (\bm{p}^{(m+1)}, t) 
 e^{i \bm{k} \cdot \bm{q}^{(m)}}, 
\end{equation}
where the sum is taken over $\bm{k}$ satisfying $\bm{k}_0 \neq \bm{0}$.

The transformed Hamiltonian is 
\begin{eqnarray}
 &&
 H^{(m+1)} = H^{(m)} + {\partial S_m \over \partial t}
\nonumber \\
 &&
\quad \quad
  = {2 \over \epsilon} (\bm{\omega} \cdot \bm{p}^{(m+1)})^{1 / 2}
    + {1 \over  \epsilon} 
      {1 \over (\bm{\omega} \cdot \bm{p}^{(m+1)})^{1 / 2}} 
       \bm{\omega} \cdot 
            {\partial S_m (\bm{q}^{(m)}, \bm{p}^{(m+1)}, t) 
             \over \partial \bm{q}^{(m)} }
     + R_1 
\nonumber \\
 &&
 \quad \quad
  + A_m (\bm{q}_1^{(m+1)}, \bm{p}^{(m+1)}, t)
  + R_2
\nonumber \\
 &&
\quad \quad
 + B_m (\bm{q}^{(m)}, \bm{p}^{(m+1)}, t)
 + R_3 
 + {\partial S_m (\bm{q}^{(m)}, \bm{p}^{(m+1)}, t)
   \over \partial t }
 \end{eqnarray}
where $R_1$, $\cdot \cdot \cdot$ $R_3$ are defined as 
\begin{eqnarray}
 &&
 R_1 = {2 \over \epsilon} (\bm{\omega} \cdot \bm{p}^{(m)})^{1 / 2}
     - {2 \over \epsilon} (\bm{\omega} \cdot \bm{p}^{(m+1)})^{1 / 2}
       - {1 \over \epsilon} 
         {\bm{\omega} \over (\bm{\omega} \cdot \bm{p}^{(m+1)})^{1 / 2}} 
         \cdot (\bm{p}^{(m)}-\bm{p}^{(m+1)}),
\\
 &&
 R_2 = A_m (\bm{q}_1^{(m)}, \bm{p}^{(m)}, t) 
     - A_m (\bm{q}_1^{(m+1)}, \bm{p}^{(m+1)}, t),
\\
 &&
 R_3 = B_m (\bm{q}^{(m)}, \bm{p}^{(m)}, t) 
     - B_m (\bm{q}^{(m)}, \bm{p}^{(m+1)}, t).
\end{eqnarray}
We determine the generating function $S_m$ so that 
the leading term depending on the fast angle variables 
$\bm{q}_0^{(m)}$ can be eliminated;
\begin{equation}
    {1 \over \epsilon} 
    {1 \over (\bm{\omega} \cdot \bm{p}^{(m+1)})^{1 / 2}} 
    \bm{\omega} \cdot {\partial S_m (\bm{q}^{(m)}, \bm{p}^{(m+1)}, t) 
               \over \partial \bm{q}^{(m)} } =
    - B_m (\bm{q}^{(m)}, \bm{p}^{(m+1)}, t).
\end{equation}
By comparing the Fourier components in the both hand sides,  
we obtain 
\begin{equation}
 S_{\bm{k}} (\bm{p}^{(m+1)}, t) = i \epsilon 
 (\bm{\omega} \cdot \bm{p}^{(m+1)})^{1 / 2}
 {1 \over (\bm{\omega} \cdot \bm{k})} b_{\bm{k}} (\bm{p}^{(m+1)}, t).
\label{AA}
\end{equation}
Before we prove the analyticity and evaluate the upper bound
of $S_m (\bm{q}^{(m)}, \bm{p}^{(m+1)}, t)$, we use the following 
lemma. 

{\it Lemma}

As for the Fourier series 
\begin{equation}
 F(\bm{q}) = \sum_{\bm{k}} F_{\bm{k}} e^{i \bm{k} \cdot \bm{q}},
\end{equation}
where 
\begin{equation}
 \bm{k} \cdot \bm{q} = \sum_{i=1}^n k_i q_i.
\end{equation}

(1) If $F(\bm{q})$ is analytic and satisfies
\begin{equation}
 |F(\bm{q})| \le C,
\end{equation}
on the domain $|{\rm Im} \bm{q}| \le \rho$, then
\begin{equation}
 | F_{\bm{k}} | \le C e^{- |\bm{k}| \rho},
\end{equation}
where
\begin{equation}
 |\bm{k}| = |k_1| + \cdot \cdot \cdot + |k_n|.
\end{equation}

(2) If on the domain $|{\rm Im} \bm{q}| \le \rho$, $F_{\bm{k}}$
satrisfies
\begin{equation}
 | F_{\bm{k}} | \le C e^{- |\bm{k}| \rho},
\end{equation}
then $F(\bm{q})$ is analytic on the domain 
$|{\rm Im} \bm{q}| \le \rho$, and for an arbitrary $\delta$
$( 0 < \delta < \rho)$, on the domain 
$|{\rm Im} \bm{q}| \le \rho - \delta$
\begin{equation}
 | F(\bm{q}) | \le { 4^n C \over \delta^n },
\end{equation}
where we assume $\delta \le 3$.

{\it Proof of Lemma} 

(1) The Fourier coefficients $F_{\bm{k}}$ are given by
\begin{equation}
 F_{\bm{k}} = {1 \over (2 \pi)^n} 
    \int^{2 \pi}_0 \cdot \cdot \cdot \int^{2 \pi}_0 
    d^n \bm{q} e^{- i \bm{k} \cdot \bm{q}} F(\bm{q}).
\end{equation}
By Cauchy's theorem, the path of integration in the above
integral can be shifted to $q_i = x_i \pm i \rho$,
$0 \le x_i \le 2 \pi$ where we choose the sign equal to
$- k_i$, we get
\begin{equation}
 |F_{\bm{k}}| \le C e^{-|\bm{k}| \rho}.
\end{equation}

(2) For an arbitrary positive $\delta$, 
$(0 < \delta <\rho)$, on the domain 
$|{\rm Im} \bm{q}| \le \rho - \delta$,
\begin{equation}
 |e^{i \bm{k} \cdot \bm{q}}| \le e^{ |\bm{k}| (\rho-\delta) }.
\end{equation}
On the domain $|{\rm Im} \bm{q}| \le \rho - \delta$
\begin{eqnarray}
  |F| &\le& \sum_{\bm{k}} |F_{\bm{k}}| 
      |e^{i \bm{k} \cdot \bm{q}}| 
\nonumber \\
      &\le& \sum_{\bm{k}} C e^{-|\bm{k}| \rho} 
                     e^{ |\bm{k}| (\rho-\delta) }
\nonumber \\
      &=& \sum_{\bm{k}} C e^{ -|\bm{k}| \delta }
\nonumber \\
      &=& C ( 1+ 2 \sum_{k > 0} e^{ -k \delta })^n
\nonumber \\
      &=& C 
     \Biggl( {1 + e^{-\delta}  \over  1 - e^{-\delta} }
     \Biggr)^n
\nonumber \\
     &<& C ( {4 \over \delta } )^n,
\end{eqnarray}
where we use the inequality
\begin{equation}
   {1 + e^{-\delta}  \over  1 - e^{-\delta} }
  < ( {4 \over \delta } ),
\end{equation}
for $0 < \delta \le 3$.
Namely on the domain 
$|{\rm Im} \bm{q}| \le \rho - \delta$, $F(\bm{q})$ converges
uniformly and absolutely, therefore $F(\bm{q})$ is analytic.
Since $\delta$ is arbitrary, $F(\bm{q})$ is analytic on 
the domain $|{\rm Im} \bm{q}| \le \rho$.

{\it Proof End of Lemma}

We evaluate the right hand side of (\ref {AA}).  
For arbitrary $\bm{p}^{(m+1)} \in D(\rho_m)$, we put
$\tau_m$ as the maximum value of $\bm{\omega} \cdot \bm{p}^{(m+1)}$;
\begin{equation}
 \sigma_m \le \bm{\omega} \cdot \bm{p}^{(m+1)}
          \le \tau_m.
\end{equation}
Since $t^m B_m$ is analytic and bounded by
$\epsilon^{m-1} M_2^{(m)}$ on the domain $D(\rho_m)$,
\begin{equation}
 |t^m b_{\bm{k}} (\bm{p}^{(m+1)}, t) | \le 
     \epsilon^{m-1} M_2^{(m)} 
     e^{- |\bm{k}| \rho_m }.
\end{equation}
As $|\bm{k}|$ increases, while $b_{\bm{k}}$ decay exponentially,
the contribution from the denominator grows like
power;
\begin{equation}
 {1 \over (\bm{\omega} \cdot \bm{k})} \le {|\bm{k}|^d \over C}.
\end{equation}
The exponential decay defeats the power grow.
We use the inequality
\begin{equation}
 |\bm{k}|^d \le e^{|\bm{k}| \delta}
 ({d \over e})^d 
 {1 \over \delta^d},
\end{equation}
for an arbitrary positive $\delta$.
The proof is as follows. 
$f(x)=x-d \ln{x}$ has a minimum at $x=d$. 
Therefore 
\begin{equation}
 { e^x \over x^d } \ge { e^d \over d^d }.
\end{equation}
For $x=|\bm{k}| \delta$, we obtain this inequality.
We evaluate the Fourier components $S_{\bm{k}}$ as
\begin{eqnarray}
 |t^m S_{\bm{k}} (\bm{p}^{(m+1)}, t)| &\le&
 \epsilon \tau_m^{1/2}
 {|\bm{k}|^d \over C} \epsilon^{m-1} M^{(m)}_2 
                        \exp( - |\bm{k}| \rho_m )
\nonumber \\
 &\le& 
 \epsilon^m \tau_m^{1/2} 
 {M^{(m)}_2 \over C} ({d \over e})^d 
 {1 \over \delta^d}
 \exp \{ - |\bm{k}| (\rho_m -\delta) \}.
\end{eqnarray}
Therefore $t^m S_m$ is analytic on the domain 
$D(\rho_m)$ and bounded as
\begin{equation}
  |t^m S_m| \le \epsilon^m L^{(m)}_1,
\end{equation}
where 
\begin{equation}
  L^{(m)}_1 = \tau_m^{1 / 2} 
 { M^{(m)}_2 \over C }
 ({d \over e})^d {4^n \over \delta^{n+d} }, 
\end{equation}
on the domain $D(\rho_m - 2 \delta)$ for an arbitrary
positive $\delta \le 3$.
We put $\rho_{m+1} = \rho_m - 3 \delta$ and on the 
domain $D(\rho_{m+1})$
\begin{eqnarray}
 |\bm{p}^{(m)} - \bm{p}^{(m+1)}| &\le& 
 |{\partial S_m \over \partial \bm{q}^{(m)} }| \le
 \epsilon^m {L^{(m)}_1 \over \delta} 
 (1 + \rho_{m+1})^m, \\
 |\bm{q}^{(m+1)} - \bm{q}^{(m)}| &\le& 
 |{\partial S_m \over \partial \bm{p}^{(m+1)} }| \le
 \epsilon^m {L^{(m)}_1 \over \delta} 
 (1 + \rho_{m+1})^m.
\end{eqnarray}
Therefore if $\epsilon$ satisfies
\begin{equation}
 \epsilon^m {L^{(m)}_1 \over \delta} 
 (1 + \rho_{m+1})^m \le 3 \delta,
\end{equation}
when $(\bm{p}^{(m+1)}, \bm{q}^{(m+1)}) \in D(\rho_{m+1})$,
$(\bm{p}^{(m)}, \bm{q}^{(m)}) \in D(\rho_{m})$.

Next on the domain $D(\rho_{m+1})$, we evaluate the residual 
part $R$ defined as 
\begin{equation}
  R = R_1 + R_2 + R_3 +
 {\partial S_m (\bm{q}^{(m)}, \bm{p}^{(m+1)}, t)
  \over \partial t }.
\end{equation}
For evaluation, we use the Taylor expansion;
\begin{equation}
 |F(1)-F(0)-\sum_{k=1}^{n-1} {1 \over k!} F^{(k)} (0)|
 \le {1 \over n!} [F^{(n)}],
\end{equation}
where $[A]$ means the maximum value of $|A|$.
As the function $F(\xi)$, we take 
\begin{equation}
 F(\xi) = f( \bm{x}_0 +\xi (\bm{x}-\bm{x}_0) ),
\end{equation}
where $\bm{x}$ is the multidimensional vector.
\begin{eqnarray}
 F^{(1)} (0) &=& {\partial f \over \partial {\bm{x}}^i } \Bigg|_0
               (\bm{x}-\bm{x}_0)^i,\\
 F^{(2)} (0) &=& 
 {\partial^2 f \over \partial \bm{x}^i \partial \bm{x}^j} \Bigg|_0
         (\bm{x}-\bm{x}_0)^i (\bm{x}-\bm{x}_0)^j.
\end{eqnarray}
The residual part is
\begin{equation}
 {1 \over n!} [F^{(n)}] =
 {1 \over n!}
         \Biggl[     
 {\partial^n F \over 
  \partial \bm{x}^{i_1} \cdot \cdot \cdot
  \partial \bm{x}^{i_n}}
   (\bm{x}-\bm{x}_0)^{i_1} \cdot \cdot \cdot 
   (\bm{x}-\bm{x}_0)^{i_n}
         \Biggr].
\end{equation}
As for $R_1$,
\begin{equation}
 |R_1| \le \Biggl[
  - {1 \over 4 \epsilon}
    {\bm{\omega}_i \bm{\omega}_j 
    \over (\bm{\omega} \cdot \bm{p})^{3/2} }
    (\bm{p}^{(m)}-\bm{p}^{(m+1)})^i
    (\bm{p}^{(m)}-\bm{p}^{(m+1)})^j
           \Biggr].
\end{equation}
Then we evaluate
\begin{equation}
  |t^{2m} R_1| \le 
         {1 \over 4 \epsilon}
         {1 \over \sigma^{3 / 2}_m}
         \Biggl\{
   \bm{\omega} n {\epsilon^m L^{(m)}_1 \over \delta} 
         \Biggr\}^2.
\end{equation}
As for $R_2$,
\begin{equation}
 |R_2| \le
            \Biggl[
 {\partial A_m \over \partial \bm{q}_1} \cdot
 (\bm{q}^{(m)}_1 -\bm{q}^{(m+1)}_1) +
 {\partial A_m \over \partial \bm{p}} \cdot
 (\bm{p}^{(m)} -\bm{p}^{(m+1)})
            \Biggr].
\end{equation}
Then we evaluate 
\begin{equation}
 |t^{m + 1} R_2| \le  
 2 n 
 {M^{(m)}_1 \over \delta}
 \epsilon^m
 {L^{(m)}_1 \over \delta}.
\end{equation}
As for $R_3$,
\begin{equation}
 |R_3| \le
          \Biggl[  
 {\partial B_m (\bm{q}^{(m)}, \bm{p}, t) 
           \over \partial \bm{p}} \cdot
 (\bm{p}^{(m)} -\bm{p}^{(m+1)})
          \Biggr].
\end{equation}
Then we evaluate
\begin{equation}
 |t^{2m} R_3| \le  
 n \epsilon^{m-1} 
 {M^{(m)}_2 \over \delta}
 \epsilon^m
 {L^{(m)}_1 \over \delta}.
\end{equation}
In addition, 
\begin{equation}
 |t^{m + 1} {\partial S_m \over \partial t}| \le
 \epsilon^m L^{(m)}_1
          (
  m + {1 + \rho_{m+1} \over \delta} 
          ).
\end{equation}
Therefore we obtain
\begin{equation}
 |t^{m+1} R| \le \epsilon^m M,
\end{equation}
where 
\begin{eqnarray}
 M &=& \epsilon^{m-1} (1 + \rho_{m+1})^{m-1} 
 {1 \over 4 \sigma^{3 / 2}_m}
 \Biggl[ \bm{\omega} n {L^{(m)}_1 \over \delta}
 \Biggr]^2 +
 2 n { M^{(m)}_1 L^{(m)}_1 \over \delta^2 }
\nonumber \\
&& + \epsilon^{m-1} n 
    (1 + \rho_{m+1})^{m-1} 
    { M^{(m)}_2 L^{(m)}_1 \over \delta^2 }
   + L^{(m)}_1 
   (m +{1+\rho_{m+1} \over \delta}).
\end{eqnarray}
We decompose $R$ into the slowly varying part $R_s$
and the fast varying part $R_f$ as
\begin{equation}
 R = R_s + R_f,
\end{equation}
where 
\begin{eqnarray}
 R_s &=& \sum_{\bm{k}_0 = \bm{0}} 
   R_{\bm{0} \bm{k}_1}(\bm{p}^{(m+1)}, t) 
   \exp{[i \bm{k}_1 \cdot \bm{q}_1^{(m+1)} ]},\\
 R_f &=& \sum_{\bm{k}_0 \neq \bm{0}} 
   R_{\bm{k}_0 \bm{k}_1}(\bm{p}^{(m+1)}, t) 
   \exp{[i \bm{k}_0 \cdot \bm{q}_0^{(m+1)} 
       + i \bm{k}_1 \cdot \bm{q}_1^{(m+1)}
        ]}.
\end{eqnarray}
Since $R_s$ is written as
\begin{equation}
 R_s = {1 \over (2 \pi)^{n_0}}
   \int^{2 \pi}_0 \cdot \cdot \cdot 
   \int^{2 \pi}_0 
   d^{n_0} \bm{q}_0^{(m+1)} R,
\end{equation}
$R_s$ is bounded as
\begin{equation}
 |t^{m+1} R_s | \le |t^{m+1} R| \le \epsilon^m M.
\end{equation}
Then 
\begin{equation}
 |t^{m+1} R_f | \le |t^{m+1} (R-R_s)| 
                \le 2 \epsilon^m M.
\end{equation}
We define 
\begin{equation}
 A_{m+1} (\bm{q}^{(m+1)}_1, \bm{p}^{(m+1)}, t)
 = A_m (\bm{q}^{(m+1)}_1, \bm{p}^{(m+1)}, t)
 + R_s (\bm{q}^{(m+1)}_1, \bm{p}^{(m+1)}, t),
\end{equation}
and
\begin{equation}
 B_{m+1} (\bm{q}^{(m+1)}, \bm{p}^{(m+1)}, t)
 = R_f (\bm{q}^{(m+1)}, \bm{p}^{(m+1)}, t).
\end{equation}
We evaluate 
\begin{eqnarray}
 &&  |t A_{m+1}| \le M_1^{(m+1)},\\
 &&  | t^{m+1} \{
      A_{m+1} (\bm{q}^{(m+1)}_1, \bm{p}^{(m+1)}, t)
    - A_m (\bm{q}^{(m+1)}_1, \bm{p}^{(m+1)}, t)
               \} |
     \le \epsilon^m {M^{(m+1)}_2 \over 2},\\
 &&  |t^{m+1} B_{m+1}| \le \epsilon^m M_2^{(m+1)},
\end{eqnarray}
where 
\begin{eqnarray}
 M_1^{(m+1)} &=& M_1^{(m)} 
     +(1+ \rho_{m+1})^m \epsilon^m M,\\
 M_2^{(m+1)} &=& 2 M,
\end{eqnarray}
on the domain $D(\rho_{m+1})$.
We complete the proof.

\section{Appendix; Proof of Propositions $4.2A$, $4.2B$}

\subsection{Proof of the Proposition $4.2A$}

We consider the system obtained by three times of canonical
transformations.
\begin{equation}
 H^{(1)} \stackrel{S_1}{\to} 
 H^{(2)} \stackrel{S_2}{\to}
 H^{(3)} \stackrel{S_3}{\to}
 H^{(4)}.  
\end{equation} 
We evaluate in how good accuracy the truncated system
$H^{(4)}_{tr}$ approximates $H^{(4)}$.
\begin{eqnarray}
 H^{(4)}_{tr} &=& {2 \over \epsilon} (\bm{\omega} \cdot \bm{P})^{1/2}
                + A_4 (\bm{Q}_1, \bm{P}, t),\\
 H^{(4)} &=& {2 \over \epsilon} (\bm{\omega} \cdot \bm{p})^{1/2}
                + A_4 (\bm{q}_1, \bm{p}, t)
                + B_4 (\bm{q}, \bm{p}, t),
\end{eqnarray}
where 
\begin{equation}
 |t A_4| \le M^{(4)}_1, \quad 
 |t^4 B_4| \le \epsilon^3 M^{(4)}_2.
\end{equation}
By taking the difference between 
\begin{equation}
 {d \bm{p}_0 \over d t} = 
 - {\partial B_4 \over \partial \bm{q}_0},
\end{equation}
and
\begin{equation}
 {d \bm{P}_0 \over d t} = 0, 
\end{equation}
we obtain
\begin{equation}
 {d \over d t} \Delta \bm{P}_0 = 
 - {\partial B_4 \over \partial \bm{q}_0}.
\end{equation}
By integrating the above inequality, we obtain
\begin{eqnarray}
 |\Delta \bm{P}_0 (t)| &\le&
 |\Delta \bm{P}_0 (1)| + \int_1 dt {\epsilon^3 \over t^4}
\nonumber \\
 &\le& 
 |\Delta \bm{P}_0 (1)| + \epsilon^3.
\end{eqnarray}
If we fix
\begin{equation}
 |\Delta \bm{P}_0 (1)| \le \epsilon^3,
\end{equation}
we obtain 
\begin{equation}
 |\Delta \bm{P}_0| \le \epsilon^3
\label{AAF}
\end{equation}

Next we evalutate $\Delta \bm{Q}_1 = \bm{q}_1 - \bm{Q}_1$.
By taking the difference between
\begin{equation}
 {d \bm{q}_1 \over d t} = 
 {\bm{\omega}_1 \over \epsilon} 
 {1 \over (\bm{\omega} \cdot \bm{p})^{1/2}} +
 {\partial A_4 \over \partial \bm{p}_1} +
 {\partial B_4 \over \partial \bm{p}_1},
\end{equation}
and
\begin{equation}
 {d \bm{Q}_1 \over d t} = 
 {\bm{\omega}_1 \over \epsilon} 
 {1 \over (\bm{\omega} \cdot \bm{P})^{1/2}} +
 {\partial A_4 \over \partial \bm{P}_1},
\end{equation}
we obtain
\begin{equation}
 {d \over d t} \Delta \bm{Q}_1 =
 {\bm{\omega}_1 \over \epsilon} 
 [ {1 \over  (\bm{\omega} \cdot \bm{p})^{1/2} } -
   {1 \over  (\bm{\omega} \cdot \bm{P})^{1/2} } ]+
 [ {\partial A_4 \over \partial \bm{p}_1} -
   {\partial A_4 \over \partial \bm{P}_1} ] +
  {\partial B_4 \over \partial \bm{p}_1}.
\end{equation}
By using the mean value theorem,
\begin{eqnarray}
&&
   {1 \over  (\bm{\omega} \cdot \bm{p})^{1/2} } -
   {1 \over  (\bm{\omega} \cdot \bm{P})^{1/2} } =
 - {1 \over 2}
   {1 \over  (\bm{\omega} \cdot \bm{P})^{3/2} }
   (\bm{\omega}_0 \cdot \Delta \bm{P}_0 
  + \bm{\omega}_1 \cdot \Delta \bm{P}_1 ),\\
&&
  {\partial A_4 \over \partial \bm{p}_1} -
  {\partial A_4 \over \partial \bm{P}_1} =
  [\Delta \bm{P}_0 \cdot {\partial \over \partial \bm{P}_0}
  +\Delta \bm{Q}_1 \cdot {\partial \over \partial \bm{Q}_1}
  +\Delta \bm{P}_1 \cdot {\partial \over \partial \bm{P}_1}]
  {\partial A_4 \over \partial \bm{P}_1},
\end{eqnarray} 
are obtained where the differentiations in the right hand 
side are taken at appropriate values between 
$(\bm{p}_0, \bm{z})$ and $(\bm{P}_0, \bm{Z})$. 
Since we are only interested in the upper bounds of the 
coefficients, we understand that the differentiations 
are taken at appropriate values between 
the original variables $(\bm{p}_0, \bm{z})$ and 
the truncated variables $(\bm{P}_0, \bm{Z})$ without notice
from now on.
Then we get
\begin{eqnarray}
 {d \over d t} \Delta \bm{Q}_1 &=&
 - {1 \over 2}
   {\bm{\omega}_1 \over \epsilon}
   {1 \over  (\bm{\omega} \cdot \bm{P})^{3/2} }
   (\bm{\omega}_0 \cdot \Delta \bm{P}_0 
  + \bm{\omega}_1 \cdot \Delta \bm{P}_1 )
\nonumber \\
  && +
    \Biggl[
   \Delta \bm{P}_0 \cdot {\partial \over \partial \bm{P}_0}
  +\Delta \bm{Q}_1 \cdot {\partial \over \partial \bm{Q}_1}
  +\Delta \bm{P}_1 \cdot {\partial \over \partial \bm{P}_1}
    \Biggr]
 {\partial A_4 \over \partial \bm{P}_1}
 +{\partial B_4 \over \partial \bm{p}_1}.
\end{eqnarray}
In the same way, as for $\Delta \bm{P}_1 = \bm{p}_1 - \bm{P}_1$,
we obtain 
\begin{equation}
  {d \over d t} \Delta \bm{P}_1 =
  -   \Biggl[
     \Delta \bm{P}_0 \cdot {\partial \over \partial \bm{P}_0}
    +\Delta \bm{Q}_1 \cdot {\partial \over \partial \bm{Q}_1}
    +\Delta \bm{P}_1 \cdot {\partial \over \partial \bm{P}_1}
      \Biggr]
   {\partial A_4 \over \partial \bm{Q}_1}
 - {\partial B_4 \over \partial \bm{q}_1}.
\end{equation}
By using the notation $\bm{z}=(\bm{q}_1, \bm{p}_1)$, 
                      $\bm{Z}=(\bm{Q}_1, \bm{P}_1)$,
$\Delta \bm{Z} = \bm{z} - \bm{Z}$, we obtain
\begin{equation}
 {d \over d t} |\Delta \bm{Z}| \le
 ({\bm{\omega}_1^2 \over \epsilon} + {\Gamma \over t}) |\Delta \bm{Z}|
+({\bm{\omega}_1 \over \epsilon} + {1 \over t}) |\Delta \bm{P}_0|
+ {\epsilon^3 \over t^4},
\end{equation}
Under the assumption $\bm{\omega}_1 \sim \epsilon$, we get
\begin{equation}
 {d \over d t} |\Delta \bm{Z}| \le
 ({\Gamma \over t}+\epsilon) |\Delta \bm{Z}|
+ |\Delta \bm{P}_0|
+ {\epsilon^3 \over t^4}.
\end{equation}
By integrating the above inequality, we obtain
\begin{equation}
 |\Delta \bm{Z}| \le 
 \exp{[ \int_1 dt ({\Gamma \over t}+\epsilon) ]}
 [ |\Delta \bm{Z} (1)| + \int_1 dt |\Delta \bm{P}_0| 
   + \int_1 dt {\epsilon^3 \over t^4}
 ].
\end{equation}
By using (\ref {AAF}), and assuming 
\begin{equation}
 |\Delta \bm{Z}(1)| \le \epsilon^3,
\end{equation}
we obtain
\begin{equation}
 |\Delta \bm{Z}| \le t^{\Gamma} \exp{[\epsilon (t-1)]}
 [\epsilon^3 t + \epsilon^3].
\end{equation}
We consider the time interval as
\begin{equation}
 1 \le t \le {1 \over t^{\beta} }
\end{equation}
where 
\begin{equation}
 \beta = {1 \over \Gamma+1} < 1.
\end{equation}
In this interval, we obtain the evaluation as
\begin{equation}
 |\Delta \bm{Z}| \le \epsilon^2.
\label{AAG}
\end{equation}

The equation for $\Delta \bm{Q}_0 = \bm{q}_0 - \bm{Q}_0$
is given by
\begin{eqnarray}
 {d \over d t} \Delta \bm{Q}_0 &=&
 - {1 \over 2}
   {\bm{\omega}_0 \over \epsilon}
   {1 \over  (\bm{\omega} \cdot \bm{P})^{3/2} }
   (\bm{\omega}_0 \cdot \Delta \bm{P}_0 
  + \bm{\omega}_1 \cdot \Delta \bm{P}_1 )
\nonumber \\
 && +
  [\Delta \bm{P}_0 \cdot {\partial \over \partial \bm{P}_0}
  +\Delta \bm{Q}_1 \cdot {\partial \over \partial \bm{Q}_1}
  +\Delta \bm{P}_1 \cdot {\partial \over \partial \bm{P}_1}]
 {\partial A_4 \over \partial \bm{P}_0}
    +
 {\partial B_4 \over \partial \bm{p}_0},
\end{eqnarray}
which yields
\begin{equation}
 \Biggl| {d \over d t} \Delta \bm{Q}_0 \Biggr| \le
 ({1 \over \epsilon} + {1 \over t})
 |\Delta \bm{P}_0| +
 {1 \over t} |\Delta \bm{Q}_1| +
 (1 + {1 \over t})
 |\Delta \bm{P}_1| +
 {\epsilon^3 \over t^4}.
\end{equation}
By integrating the above inequality, we obtain
\begin{equation}
 |\Delta \bm{Q}_0| \le |\Delta \bm{Q}_0 (1)| +
 {1 \over \epsilon} \int_1 dt |\Delta \bm{P}_0|+ 
 \int_1 {d t \over t} |\Delta \bm{Q}_1|+
 \int_1 dt |\Delta \bm{P}_1|+
 \int_1 dt {\epsilon^3 \over t^4},
\end{equation}
which yields
\begin{eqnarray}
 \Biggl|{\Delta \bm{Q}_0 \over t}
 \Biggr| &\le&
 | \Delta \bm{Q}_0 (1) | +
 {1 \over \epsilon} |\Delta \bm{P}_0|+
 |\Delta \bm{Z}| +\epsilon^3 
\nonumber \\
 &\le& \epsilon^2,
\label{AAGA}
\end{eqnarray}
where we used (\ref {AAF}), (\ref {AAG}) and assumed
that
\begin{equation}
 |\Delta \bm{Q}_0 (1)| \le \epsilon^2.
\end{equation}
By obtaining (\ref {AAF}), (\ref {AAG}), (\ref {AAGA}),
we complete the proof of (\ref {AAA}).

Next we consider in how good accuracy the perturbations
of the truncated system $H^{(4)}_{tr}$ approximate 
the perturbations of the transformed system $H^{(4)}$.
We try to obtain the equation for 
$\Delta \delta \bm{Q}_0 = \delta \bm{q}_0 - \delta \bm{Q}_0$.
We take the difference between 
\begin{eqnarray}
 {d \over d t} \delta \bm{q}_0 &=&
  - {1 \over 2}
   {\bm{\omega}_0 \over \epsilon} 
   {1 \over  (\bm{\omega} \cdot \bm{p})^{3/2} }
   (\bm{\omega}_0 \cdot \delta \bm{p}_0 
  + \bm{\omega}_1 \cdot \delta \bm{p}_1 )
\nonumber \\
 && +
  (\delta \bm{p}_0 \cdot {\partial \over \partial \bm{p}_0}
  +\delta \bm{q}_1 \cdot {\partial \over \partial \bm{q}_1}
  +\delta \bm{p}_1 \cdot {\partial \over \partial \bm{p}_1})
  {\partial A_4 \over \partial \bm{p}_0}
\nonumber \\
 && +
  (\delta \bm{q}_0 \cdot {\partial \over \partial \bm{q}_0}
  +\delta \bm{p}_0 \cdot {\partial \over \partial \bm{p}_0}
  +\delta \bm{q}_1 \cdot {\partial \over \partial \bm{q}_1}
  +\delta \bm{p}_1 \cdot {\partial \over \partial \bm{p}_1})
  {\partial B_4 \over \partial \bm{p}_0},
\end{eqnarray}
and 
\begin{equation}
  {d \over d t} \delta \bm{Q}_0 =
  - {1 \over 2}
   {\bm{\omega}_0 \over \epsilon} 
   {1 \over  (\bm{\omega} \cdot \bm{P})^{3/2} }
   (\bm{\omega}_0 \cdot \delta \bm{P}_0 
  + \bm{\omega}_1 \cdot \delta \bm{P}_1 )
   +
  (\delta \bm{P}_0 \cdot {\partial \over \partial \bm{P}_0}
  +\delta \bm{Q}_1 \cdot {\partial \over \partial \bm{Q}_1}
  +\delta \bm{P}_1 \cdot {\partial \over \partial \bm{P}_1})
  {\partial A_4 \over \partial \bm{P}_0}.
\end{equation}
By using the mean value theorem, we obtain
\begin{eqnarray}
 &&
 - {1 \over 2}
   {\bm{\omega}_0 \over \epsilon}
   {1 \over  (\bm{\omega} \cdot \bm{p})^{3/2} }
   (\bm{\omega}_0 \cdot \delta \bm{p}_0 
  + \bm{\omega}_1 \cdot \delta \bm{p}_1 )
 + {1 \over 2}
   {\bm{\omega}_0 \over \epsilon}
   {1 \over  (\bm{\omega} \cdot \bm{P})^{3/2} }
   (\bm{\omega}_0 \cdot \delta \bm{P}_0 
  + \bm{\omega}_1 \cdot \delta \bm{P}_1 )
\nonumber \\
&& =
 - {1 \over 2}
   {\bm{\omega}_0 \over \epsilon}
   {1 \over  (\bm{\omega} \cdot \bm{p})^{3/2} }
   (\bm{\omega}_0 \cdot \Delta \delta \bm{P}_0 
  + \bm{\omega}_1 \cdot \Delta \delta \bm{P}_1 )
\nonumber \\
&& \quad \quad 
 + {3 \over 4}
   {\bm{\omega}_0 \over \epsilon}
   {1 \over  (\bm{\omega} \cdot \bm{P})^{5/2} }
   (\bm{\omega}_0 \cdot \Delta \bm{P}_0 
  + \bm{\omega}_1 \cdot \Delta \bm{P}_1 )
   (\bm{\omega}_0 \cdot \delta \bm{P}_0 
  + \bm{\omega}_1 \cdot \delta \bm{P}_1 ),
\end{eqnarray}
\begin{eqnarray}
 &&
  (\delta \bm{p}_0 \cdot {\partial \over \partial \bm{p}_0}
  +\delta \bm{q}_1 \cdot {\partial \over \partial \bm{q}_1}
  +\delta \bm{p}_1 \cdot {\partial \over \partial \bm{p}_1})
 {\partial A_4 \over \partial \bm{p}_0}
 -
  (\delta \bm{P}_0 \cdot {\partial \over \partial \bm{P}_0}
  +\delta \bm{Q}_1 \cdot {\partial \over \partial \bm{Q}_1}
  +\delta \bm{P}_1 \cdot {\partial \over \partial \bm{P}_1})
 {\partial A_4 \over \partial \bm{P}_0}
\nonumber \\
&& =
  (\Delta \delta \bm{P}_0 \cdot {\partial \over \partial \bm{p}_0}
  +\Delta \delta \bm{Q}_1 \cdot {\partial \over \partial \bm{q}_1}
  +\Delta \delta \bm{P}_1 \cdot {\partial \over \partial \bm{p}_1})
 {\partial A_4 \over \partial \bm{p}_0}
\nonumber \\
&& +
  (\Delta \bm{P}_0 \cdot {\partial \over \partial \bm{P}_0}
  +\Delta \bm{Q}_1 \cdot {\partial \over \partial \bm{Q}_1}
  +\Delta \bm{P}_1 \cdot {\partial \over \partial \bm{P}_1})
\nonumber \\
&& \quad \quad
  (\delta \bm{P}_0 \cdot {\partial \over \partial \bm{P}_0}
  +\delta \bm{Q}_1 \cdot {\partial \over \partial \bm{Q}_1}
  +\delta \bm{P}_1 \cdot {\partial \over \partial \bm{P}_1})
  {\partial A_4 \over \partial \bm{P}_0},
\end{eqnarray}
and 
\begin{eqnarray}
 &&
  (\delta \bm{q}_0 \cdot {\partial \over \partial \bm{q}_0}
  +\delta \bm{p}_0 \cdot {\partial \over \partial \bm{p}_0}
  +\delta \bm{q}_1 \cdot {\partial \over \partial \bm{q}_1}
  +\delta \bm{p}_1 \cdot {\partial \over \partial \bm{p}_1})
  {\partial B_4 \over \partial \bm{p}_0}
\nonumber \\
 && =
  (\delta \bm{Q}_0 \cdot {\partial \over \partial \bm{q}_0}
  +\delta \bm{P}_0 \cdot {\partial \over \partial \bm{p}_0}
  +\delta \bm{Q}_1 \cdot {\partial \over \partial \bm{q}_1}
  +\delta \bm{P}_1 \cdot {\partial \over \partial \bm{p}_1})
  {\partial B_4 \over \partial \bm{p}_0}
\nonumber \\
 && \quad \quad +
  (\Delta \delta \bm{Q}_0 \cdot {\partial \over \partial \bm{q}_0}
  +\Delta \delta \bm{P}_0 \cdot {\partial \over \partial \bm{p}_0}
  +\Delta \delta \bm{Q}_1 \cdot {\partial \over \partial \bm{q}_1}
  +\Delta \delta \bm{P}_1 \cdot {\partial \over \partial \bm{p}_1})
  {\partial B_4 \over \partial \bm{p}_0},
\end{eqnarray}
where the differentiations in the right hand side 
are taken at the appropriate values between 
$(\bm{Q}, \bm{P})$ and $(\bm{q}, \bm{p})$, and it will not be noticed
from now on.
By evaluating the coefficients, we obtain
\begin{eqnarray}
 \Biggl| {d \over d t} \Delta \delta \bm{Q}_0 \Biggr|
 &\le& 
 {1 \over \epsilon} 
 (|\Delta \delta \bm{P}_0|+\bm{\omega}_1 |\Delta \delta \bm{P}_1|)
 +
 {1 \over \epsilon} 
 (|\Delta \bm{P}_0|+\bm{\omega}_1 |\Delta \bm{P}_1|)
 (|\delta \bm{P}_0|+\bm{\omega}_1 |\delta \bm{P}_1|)
\nonumber \\
&& +
 (|\Delta \delta \bm{P}_0|+ |\Delta \delta \bm{Q}_1|
 +|\Delta \delta \bm{P}_1|) {1 \over t}
\nonumber \\
&& +
 (|\Delta \bm{P}_0|+ |\Delta \bm{Q}_1|+|\Delta \bm{P}_1|) 
 (|\delta \bm{P}_0|+ |\delta \bm{Q}_1|+|\delta \bm{P}_1|) 
{1 \over t}
\nonumber \\
&& +
 (|\delta \bm{Q}_0|+ |\delta \bm{P}_0|+ |\delta \bm{Q}_1|
 +|\delta \bm{P}_1|) {\epsilon^3 \over t^4}
\nonumber \\
&& +
 (|\Delta \delta \bm{Q}_0|+ |\Delta \delta \bm{P}_0|+ 
  |\Delta \delta \bm{Q}_1|+ |\Delta \delta \bm{P}_1|) 
 {\epsilon^3  \over t^4}.
\end{eqnarray}
By using the inequalities as
\begin{eqnarray}
 && 
|\Delta \bm{P}_0|+ |\Delta \bm{Q}_1|+|\Delta \bm{P}_1| 
 \le \epsilon^2,\\
 &&
 |\Delta \bm{P}_0|+\bm{\omega}_1 |\Delta \bm{P}_1|
 \le \epsilon^3,
\end{eqnarray}
we obtain 
\begin{eqnarray}
 \Biggl| {d \over d t} \Delta \delta \bm{Q}_0
 \Biggr|
 &\le&
 ({1 \over \epsilon}+{1 \over t})|\Delta \delta \bm{P}_0|
 + |\Delta \delta \bm{P}_1|
 + {1 \over t} |\Delta \delta \bm{Q}_1|
 + {\epsilon^3 \over t^4} |\Delta \delta \bm{Q}_0|
\nonumber \\
&& 
 + \epsilon^2 |\delta \bm{P}_0|
 + \epsilon^2 (\epsilon +{1 \over t})|\delta \bm{P}_1|
 + {\epsilon^2 \over t} |\delta \bm{Q}_1|
 + {\epsilon^3 \over t^4} |\delta \bm{Q}_0|.
\end{eqnarray}
In the same way as $\Delta \delta \bm{Q}_0$, we obtain
\begin{eqnarray}
 \Biggl| {d \over d t} \Delta \delta \bm{P}_0
 \Biggr| &\le&
 (|\Delta \delta \bm{Q}_0|+ |\Delta \delta \bm{P}_0|+ 
  |\Delta \delta \bm{Q}_1|+ |\Delta \delta \bm{P}_1|) 
 {\epsilon^3 \over t^4}
\nonumber \\
 && +
 (|\delta \bm{Q}_0|+ |\delta \bm{P}_0|+ 
  |\delta \bm{Q}_1|+ |\delta \bm{P}_1|) 
 {\epsilon^3 \over t^4},
\end{eqnarray}
and
\begin{eqnarray}
&&
  {d \over d t}
   |\Delta \delta \bm{Z}| \le
  \left( \frac{\Gamma}{t} + \epsilon  
  \right)
   |\Delta \delta \bm{Z}|
+ |\Delta \delta \bm{P}_0|
+ {\epsilon^3 \over t^4} |\Delta \delta \bm{Q}_0|
\nonumber \\
&& + \epsilon^2 (\epsilon + \dfrac{1}{t})   
 |\delta \bm{P}_0|
+ {\epsilon^2 \over t} |\delta \bm{Q}_1|
+  \epsilon^2 (\epsilon^2 + \dfrac{1}{t})   
  |\delta \bm{P}_1|
+ {\epsilon^3 \over t^4} |\delta \bm{Q}_0|.
\end{eqnarray}
By using 
\begin{equation}
 |\Delta \delta \bm{Q}_1|,|\Delta \delta \bm{P}_1| \le
 |\Delta \delta \bm{Z}|,
\quad
 |\delta \bm{Q}_1|,|\delta \bm{P}_1| \le
 |\delta \bm{Z}| 
\end{equation}
we obtain
\begin{eqnarray}
   {d \over d t} |\Delta \delta \bm{Z}|
 &\le&
   (\epsilon + {\Gamma \over t}) |\Delta \delta \bm{Z}|
 + |\Delta \delta \bm{P}_0|
 + {\epsilon^3 \over t^4} |\Delta \delta \bm{Q}_0|
\nonumber \\
&& 
 + \epsilon^2 |\delta \bm{P}_0|
 + \epsilon^2 |\delta \bm{Z}|
 + {\epsilon^3 \over t^4} |\delta \bm{Q}_0|,
\label{AAH}
\end{eqnarray}
\begin{eqnarray}
   \Biggl| {d \over d t} \Delta \delta \bm{Q}_0
   \Biggr|
 &\le&
   {1 \over \epsilon} |\Delta \delta \bm{P}_0|
 + |\Delta \delta \bm{Z}|
 + {\epsilon^3 \over t^4} |\Delta \delta \bm{Q}_0|
\nonumber \\
&& 
 + \epsilon^2 |\delta \bm{P}_0|
 + \epsilon^2 |\delta \bm{Z}|
 + {\epsilon^3 \over t^4} |\delta \bm{Q}_0|,
\label{AAI}
\end{eqnarray}
and
\begin{eqnarray}
   \Biggl| {d \over d t} \Delta \delta \bm{P}_0
   \Biggr|
 &\le&
 ( |\Delta \delta \bm{Q}_0|+ 
   |\Delta \delta \bm{P}_0|+
   |\Delta \delta \bm{Z}| )
   {\epsilon^3 \over t^4} 
\nonumber \\
&& +
 ( |\delta \bm{Q}_0|+ |\delta \bm{P}_0|+ |\delta \bm{Z}| )
   {\epsilon^3 \over t^4}. 
\label{AAJ}
\end{eqnarray}
The inequality (\ref {AAH}) yields
\begin{eqnarray}
 |\Delta \delta \bm{Z}| &\leq&
 \exp[ \int_1 dt (\epsilon + {\Gamma \over t}) ]
\nonumber \\
 && 
 \{ 
 |\Delta \delta \bm{Z} (1)| + 
 \int_1 dt |\Delta \delta \bm{P}_0| +
 \int_1 dt {\epsilon^3 \over t^4} |\Delta \delta \bm{Q}_0| 
\nonumber \\
&& +
 \int_1 dt \epsilon^2 |\delta \bm{P}_0| +
 \int_1 dt \epsilon^2 |\delta \bm{Z}| + 
 \int_1 dt {\epsilon^3 \over t^4} |\delta \bm{Q}_0| 
 \}
\nonumber \\
&\le&
 t^{\Gamma} \exp[\epsilon (t-1)]
\nonumber \\
&&
 \{
 |\Delta \delta \bm{Z} (1)| +
 t \| \Delta \delta \bm{P}_0 \| +
 \epsilon^3 \Biggl\|{
    \Delta \delta \bm{Q}_0 \over t^2}
            \Biggr\| +
 \epsilon^2 t \| \delta \bm{P}_0 \| 
\nonumber \\
&& +
 \epsilon^2 t \| \delta \bm{Z} \| +
 \epsilon^3 \Biggl\|
    {\delta \bm{Q}_0 \over t^2}
            \Biggr\|
 \}.
\end{eqnarray}
The inequality (\ref {AAI}) yields
\begin{eqnarray}
  |\Delta \delta \bm{Q}_0| &\leq&
 \exp[ \int_1 dt {\epsilon^3 \over t^4} ]
\nonumber \\
 && 
 \{ 
 |\Delta \delta \bm{Q}_0 (1)| + 
 {t \over \epsilon} \| \Delta \delta \bm{P}_0 \| +
 t \| \Delta \delta \bm{Z} \| 
\nonumber \\
&&
 +  \epsilon^2 t \| \delta \bm{P}_0 \| 
 +  \epsilon^2 t \| \delta \bm{Z} \| 
 +  \epsilon^3 \Biggl\| {\delta \bm{Q}_0 \over t^2} \Biggr\| 
 \}
\nonumber \\
&\le&
 \{ 
 |\Delta \delta \bm{Q}_0 (1)| + 
 {t \over \epsilon} \| \Delta \delta \bm{P}_0 \| +
 t \| \Delta \delta \bm{Z} \| 
\nonumber \\
&&
 +  \epsilon^2 t \| \delta \bm{P}_0 \| 
 +  \epsilon^2 t \| \delta \bm{Z} \| 
 +  \epsilon^3 \Biggl\| {\delta \bm{Q}_0 \over t^2} \Biggr\| 
 \},
\end{eqnarray}
and the inequality (\ref {AAJ}) yields
\begin{eqnarray}
 |\Delta \delta \bm{P}_0| 
 &\le&
 |\Delta \delta \bm{P}_0 (1)|
\nonumber \\
 &&
 + \epsilon^3 
 (    
   \Biggl\| {\Delta \delta \bm{Q}_0 \over t^2} \Biggr\| 
 + \Biggl\| {\Delta \delta \bm{Z} \over t^2} \Biggr\| 
 + \Biggl\| {\delta \bm{Q}_0 \over t^2} \Biggr\|
 + \Biggl\| {\delta \bm{P}_0 \over t^2} \Biggr\|  
 + \Biggl\| {\delta \bm{Z} \over t^2} \Biggr\| 
 ) .
\end{eqnarray}
If we consider the time interval
\begin{equation}
 1 \le t \le {1 \over t^{\beta}},
\label{AAJA}
\end{equation}
where
\begin{equation}
 \beta = {1 \over \Gamma +1} \le 1,
\end{equation}
the growth factor of $|\Delta \delta \bm{Z}|$ is 
bounded as 
\begin{equation}
 t^{\Gamma} \exp[\epsilon (t-1)] t \le {1 \over \epsilon}.
\end{equation}
Then in this interval (\ref {AAJA}), we obtain
\begin{eqnarray}
  \Biggl| {\Delta \delta \bm{Q}_0 \over t} \Biggr| 
 &\le&
   |\Delta \delta \bm{Q}_0 (1) |
 + {1 \over \epsilon}  \| \Delta \delta \bm{P}_0 \|
 + \| \Delta \delta \bm{Z}\|
\nonumber \\
 && 
  + \epsilon^2 \| \delta \bm{P}_0 \|
  + \epsilon^2 \| \delta \bm{Z} \|
  + \epsilon^3 \Biggl\| {\delta \bm{Q}_0 \over t^2} \Biggr\|,
\label{AAK}
\end{eqnarray}
\begin{eqnarray}
  |\Delta \delta \bm{P}_0| 
 &\le&
   |\Delta \delta \bm{P}_0 (1)|
 + \epsilon^3 (
   \Biggl\| {\Delta \delta \bm{Q}_0 \over t^2} \Biggr\|
 + \Biggl\| {\Delta \delta \bm{Z} \over t^2} \Biggr\|
\nonumber \\
 && 
  + \Biggl\| {\delta \bm{Q}_0 \over t^2} \Biggr\|
  + \Biggl\| {\delta \bm{P}_0 \over t^2} \Biggr\|
  + \Biggl\| {\delta \bm{Z} \over t^2} \Biggr\|
             ),
\label{AAL}
\end{eqnarray}
\begin{eqnarray}
  |\Delta \delta \bm{Z}| 
 &\le&
   {1 \over \epsilon} |\Delta \delta \bm{Z}(1)|
 + {1 \over \epsilon} \| \Delta \delta \bm{P}_0 \|
 + \epsilon^2 \Biggl\| {\Delta \delta \bm{Q}_0 \over t^2} \Biggr\|
\nonumber \\
 && 
  + \epsilon \| \delta \bm{P}_0 \|
  + \epsilon \| \delta \bm{Z} \|
  + \epsilon^2 \Biggl\| {\delta \bm{Q}_0 \over t^2} \Biggr\|,
\label{AAM}
\end{eqnarray}
By using (\ref {AAL}) to (\ref {AAK}) (\ref {AAM}),
we obtain
\begin{eqnarray}
  \Biggl\| {\Delta \delta \bm{Q}_0 \over t} \Biggr\| 
 &\le&
   |\Delta \delta \bm{Q}_0 (1) |
 + {1 \over \epsilon}  |\Delta \delta \bm{P}_0 (1)|
 + \| \Delta \delta \bm{Z} \|
\nonumber \\
 && 
  + \epsilon^2 \Biggl\| {\delta \bm{Q}_0 \over t^2} \Biggr\|,
  + \epsilon^2 \| \delta \bm{Z} \|
  + \epsilon^2 \| \delta \bm{P}_0 \|,
\label{AAN}
\end{eqnarray}
and 
\begin{eqnarray}
  \| \Delta \delta \bm{Z} \| 
 &\le&
   {1 \over \epsilon} |\Delta \delta \bm{Z}(1)|
 + {1 \over \epsilon} |\Delta \delta \bm{P}_0 (1)|
\nonumber \\
 && 
  + \epsilon^2 \Biggl\| {\Delta \delta \bm{Q}_0 \over t^2} \Biggr\|
  + \epsilon^2 \Biggl\| {\delta \bm{Q}_0 \over t^2} \Biggr\|
  + \epsilon \| \delta \bm{P}_0 \|
  + \epsilon \| \delta \bm{Z} \| .
\label{AAO}
\end{eqnarray}
By using (\ref {AAO}) to (\ref {AAN}), we get
\begin{eqnarray}
  \Biggl\| {\Delta \delta \bm{Q}_0 \over t} \Biggr\| 
 &\le&
   |\Delta \delta \bm{Q}_0 (1) |
 + {1 \over \epsilon}  |\Delta \delta \bm{P}_0 (1)|
 + {1 \over \epsilon}  |\Delta \delta \bm{Z} (1)|
\nonumber \\
 &&
  + \epsilon^2 \Biggl\| {\delta \bm{Q}_0 \over t^2} \Biggr\|,
  + \epsilon \| \delta \bm{P}_0 \|
  + \epsilon \| \delta \bm{Z} \|,
\label{AAP}
\end{eqnarray}
By using (\ref {AAP}) to (\ref {AAO}), we obtain
\begin{eqnarray}
  \| \Delta \delta \bm{Z} \| 
 &\le&
   \epsilon^2 |\Delta \delta \bm{Q}_0 (1)|
 + {1 \over \epsilon} |\Delta \delta \bm{P}_0 (1)|
 + {1 \over \epsilon} |\Delta \delta \bm{Z}(1)|
\nonumber \\
 && 
  + \epsilon^2 \Biggl\| {\delta \bm{Q}_0 \over t^2} \Biggr\|
  + \epsilon \| \delta \bm{P}_0 \|
  + \epsilon \| \delta \bm{Z} \| .
\label{AAQ}
\end{eqnarray}
By using (\ref {AAP})(\ref {AAQ}) to (\ref {AAL}),
we obtain
\begin{eqnarray}
 \| \Delta \delta \bm{P}_0 \| 
 &\le& 
   \epsilon^3 |\Delta \delta \bm{Q}_0 (1)|
 + |\Delta \delta \bm{P}_0 (1)|
 + \epsilon^2 |\Delta \delta \bm{Z} (1)|
\nonumber \\
 &&
 + \epsilon^3 \Biggl\| {\delta \bm{Q}_0 \over t^2} \Biggr\|
 + \epsilon^3 \| \delta \bm{P}_0 \| 
 + \epsilon^3 \| \delta \bm{Z} \| .
\label{AAQA} 
\end{eqnarray}
By substituting 
\begin{equation}
   \Delta \delta \bm{Q}_0 (1)
 = \Delta \delta \bm{P}_0 (1)
 = \Delta \delta \bm{Z} (1)
 = 0
\end{equation}
to (\ref {AAP}) (\ref {AAQ}) (\ref {AAQA}), we obtain
\begin{eqnarray}
  \Biggl\| {\Delta \delta \bm{Q}_0 \over t} \Biggr\| 
  &\le& \epsilon^2 \Biggl\| {\delta \bm{Q}_0 \over t^2} \Biggr\|
      + \epsilon \| \delta \bm{P}_0 \|
      + \epsilon \| \delta \bm{Z} \|,
\label{AAR} \\
 \| \Delta \delta \bm{P}_0 \| 
  &\le& \epsilon^3 \Biggl\| {\delta \bm{Q}_0 \over t^2} \Biggr\|
      + \epsilon^3 \| \delta \bm{P}_0 \|
      + \epsilon^3 \| \delta \bm{Z}\|,
\label{AAS} \\
 \| \Delta \delta \bm{Z} \| 
  &\le& \epsilon^2 \Biggl\| {\delta \bm{Q}_0 \over t^2} \Biggr\|
      + \epsilon \| \delta \bm{P}_0 \|
      + \epsilon \| \delta \bm{Z} \|.
\label{AAT}
\end{eqnarray}
Next we evaluate $\| \delta \bm{Q}_0 / t^2 \|$ by
\begin{eqnarray}
 {d \over d t} \delta \bm{Q}_0 
 &=&
 - {1 \over 2}
   {\bm{\omega}_0 \over \epsilon}
   {1 \over  (\bm{\omega} \cdot \bm{P})^{3/2} }
   (\bm{\omega}_0 \cdot \delta \bm{P}_0 
  + \bm{\omega}_1 \cdot \delta \bm{P}_1 )
\nonumber \\
 && +
  (\delta \bm{P}_0 \cdot {\partial \over \partial \bm{P}_0}
  +\delta \bm{Q}_1 \cdot {\partial \over \partial \bm{Q}_1}
  +\delta \bm{P}_1 \cdot {\partial \over \partial \bm{P}_1})
 {\partial A_4 \over \partial \bm{P}_0},
\end{eqnarray}
which yields
\begin{equation}
 \Biggl\| {\delta \bm{Q}_0 \over t} \Biggr\|
 \le |\delta \bm{Q}_0 (1)|
   + {1 \over \epsilon} \| \delta \bm{P}_0 \|
   + \| \delta \bm{Z} \|.
\label{AAU}
\end{equation}
By using (\ref {AAU}) 
to (\ref {AAR}) (\ref {AAS}) (\ref {AAT}), 
we get
\begin{eqnarray}
  \Biggl\| {\Delta \delta \bm{Q}_0 \over t} \Biggr\| 
  &\le& \epsilon^2 |\delta \bm{Q}_0 (1)|
      + \epsilon \| \delta \bm{P}_0 \|
      + \epsilon \| \delta \bm{Z} \|,
\\
 \| \Delta \delta \bm{P}_0 \| 
  &\le& \epsilon^3 |\delta \bm{Q}_0 (1)|
      + \epsilon^2 \| \delta \bm{P}_0 \|
      + \epsilon^3 \| \delta \bm{Z} \|,
\\
 \| \Delta \delta \bm{Z} \| 
  &\le& \epsilon^2 |\delta \bm{Q}_0 (1)|
      + \epsilon \| \delta \bm{P}_0 \|
      + \epsilon \| \delta \bm{Z} \|.
\end{eqnarray}
These complete the proof of (\ref {AAB}).

\subsection{Proof of Proposition $4.2B$}

We evaluate how the errors 
$\Delta \bm{Q}^{(4)}$, $\Delta \bm{P}^{(4)}$,
$\Delta \delta \bm{Q}^{(4)}$, $\Delta \delta \bm{P}^{(4)}$
are transmitted to
$\Delta \bm{Q}^{(1)}$, $\Delta \bm{P}^{(1)}$,
$\Delta \delta \bm{Q}^{(1)}$, $\Delta \delta \bm{P}^{(1)}$
by the canonical transformations
\begin{eqnarray}
 \bm{p}^{(m)} &=& \bm{p}^{(m+1)} 
      + {\partial S_m \over \partial \bm{q}^{(m)} },\\
 \bm{q}^{(m+1)} &=& \bm{q}^{(m)} 
      + {\partial S_m \over \partial \bm{p}^{(m+1)} },
\quad \quad (m \ge 1)
\end{eqnarray}
where 
\begin{equation}
 |t^m S_m| \le \epsilon^m L_1^{(m)}.
\end{equation}

{\it Lemma}
\begin{eqnarray}
 |\Delta \bm{P}^{(1)}-\Delta \bm{P}^{(m)}|
 &\le& {\epsilon \over t} |\Delta \bm{Q}^{(m)}|
     + {\epsilon \over t} |\Delta \bm{P}^{(m)}|,
\nonumber \\
 |\Delta \bm{Q}^{(1)}-\Delta \bm{Q}^{(m)}|
 &\le& {\epsilon \over t} |\Delta \bm{Q}^{(m)}|
     + {\epsilon \over t} |\Delta \bm{P}^{(m)}|,
\label{AAAA}
\end{eqnarray}
\begin{eqnarray}
 |\delta \bm{P}^{(1)}-\delta \bm{P}^{(m)}|
 &\le& {\epsilon \over t} |\delta \bm{Q}^{(m)}|
     + {\epsilon \over t} |\delta \bm{P}^{(m)}|,
\nonumber \\
 |\delta \bm{Q}^{(1)}-\delta \bm{Q}^{(m)}|
 &\le& {\epsilon \over t} |\delta \bm{Q}^{(m)}|
     + {\epsilon \over t} |\delta \bm{P}^{(m)}|,
\label{AAAB}
\end{eqnarray}
{\it Proof of Lemma}

We can prove (\ref {AAAA}), (\ref {AAAB}) in the almost
same way.
So we prove (\ref {AAAA}) as the representative.
We consider the difference $\Delta \bm{P} = \bm{p} - \bm{P}$.
Taking the difference between
\begin{equation}
  \bm{p}^{(m)} = \bm{p}^{(m+1)} 
      + {\partial S_m \over \partial \bm{q}^{(m)} },
\end{equation}
and
\begin{equation}
  \bm{P}^{(m)} = \bm{P}^{(m+1)} 
      + {\partial S_m \over \partial \bm{Q}^{(m)} },
\end{equation}
yields
\begin{eqnarray}
 \Delta \bm{P}^{(m)} 
 &=& 
 \Delta \bm{P}^{(m+1)} 
 + {\partial S_m \over \partial \bm{q}^{(m)} }
 - {\partial S_m \over \partial \bm{Q}^{(m)} }
\nonumber \\
 &=&
 \Delta \bm{P}^{(m+1)} 
 +(
   \Delta \bm{Q}^{(m)} \cdot {\partial \over \partial \bm{Q}^{(m)}}
  +\Delta \bm{P}^{(m+1)} \cdot {\partial \over \partial \bm{P}^{(m+1)}}
  )
 {\partial S_m \over \partial \bm{Q}^{(m)} },
\end{eqnarray}
where differentiations in the right hand side are taken
at the appropriate values between $(\bm{q}, \bm{p})$ 
and $(\bm{Q}, \bm{P})$
according the mean value theorem.
So we obtain
\begin{eqnarray}
 |\Delta \bm{P}^{(m)}-\Delta \bm{P}^{(m+1)}|
 &\le& {\epsilon^m \over t^m} 
(|\Delta \bm{Q}^{(m)}|+ |\Delta \bm{P}^{(m+1)}|),
\nonumber \\
 |\Delta \bm{Q}^{(m)}-\Delta \bm{Q}^{(m+1)}|
 &\le& {\epsilon^m \over t^m} 
(|\Delta \bm{Q}^{(m)}|+ |\Delta \bm{P}^{(m+1)}|),
\label{ABA}
\end{eqnarray}
As for the first term in the right hand side of the above
inequalities, we obtain
\begin{eqnarray}
 |\Delta \bm{Q}^{(m)}| 
 &\le&
   |\Delta \bm{Q}^{(m)}- \Delta \bm{Q}^{(m+1)}|
 + |\Delta \bm{Q}^{(m+1)}|  
\nonumber \\
 &\le&
 {\epsilon^m \over t^m} 
 (|\Delta \bm{Q}^{(m)}|+ |\Delta \bm{P}^{(m+1)}|)
 + |\Delta \bm{Q}^{(m+1)}|, 
\end{eqnarray}
which yields
\begin{equation}
 |\Delta \bm{Q}^{(m)}| \le 
 |\Delta \bm{Q}^{(m+1)}| 
 + {\epsilon^m \over t^m} |\Delta \bm{P}^{(m+1)}|.
\label{ABB} 
\end{equation}
By using (\ref {ABB}) to (\ref {ABA}), we obtain
\begin{eqnarray}
 |\Delta \bm{P}^{(m)}-\Delta \bm{P}^{(m+1)}|
&\le& 
 {\epsilon^m \over t^m} |\Delta \bm{X}^{(m+1)}|,
\nonumber \\
  |\Delta \bm{Q}^{(m)}-\Delta \bm{Q}^{(m+1)}|
&\le& 
 {\epsilon^m \over t^m} |\Delta \bm{X}^{(m+1)}|,
\label{ABC}
\end{eqnarray}
where 
\begin{equation}
 |\Delta \bm{X}^{(m)}|= 
 |\Delta \bm{Q}^{(m)}|+|\Delta \bm{P}^{(m)}|.
\end{equation}
In the same way as (\ref {ABB}), we obtain
\begin{equation}
 |\Delta \bm{P}^{(m)}|= 
 |\Delta \bm{P}^{(m+1)}|
 +{\epsilon^m \over t^m} |\Delta \bm{Q}^{(m+1)}|.
\label{ABD}
\end{equation}
From (\ref {ABB}) (\ref {ABD}), we obtain 
\begin{equation}
 |\Delta \bm{X}^{(m)}| \le |\Delta \bm{X}^{(m+1)}|.
\label{ABE}
\end{equation}
We evaluate
\begin{eqnarray}
 |\Delta \bm{P}^{(1)}- \Delta \bm{P}^{(m)}|
&\le&
  |\Delta \bm{P}^{(1)}- \Delta \bm{P}^{(2)}|
+ |\Delta \bm{P}^{(2)}- \Delta \bm{P}^{(3)}|
+ \cdot \cdot \cdot
+ |\Delta \bm{P}^{(m-1)}- \Delta \bm{P}^{(m)}|
\nonumber \\
&\le& 
 {\epsilon \over t} |\Delta \bm{X}^{(2)}|
+{\epsilon^2 \over t^2} |\Delta \bm{X}^{(3)}|
+ \cdot \cdot \cdot
+{\epsilon^{m-1} \over t^{m-1}} |\Delta \bm{X}^{(m)}|
\nonumber \\
&\le&
 {\epsilon \over t} |\Delta \bm{X}^{(m)}|
\end{eqnarray}
where we used (\ref {ABC}) (\ref {ABE}).
In the same way, we get
\begin{equation}
 |\Delta \bm{Q}^{(1)}- \Delta \bm{Q}^{(m)}|
\le {\epsilon \over t} |\Delta \bm{X}^{(m)}|.
\end{equation}
These complete the proof.

{\it Proof of Lemma}

From the inequalities (\ref {AAA}), we obtain
\begin{equation}
 {1 \over t} |\Delta \bm{Q}^{(4)}|
+{1 \over t} |\Delta \bm{P}^{(4)}|
\le \epsilon^2.
\end{equation}
So 
\begin{eqnarray}
 |\Delta \bm{P}^{(1)}-\Delta \bm{P}^{(4)}|
&\le& \epsilon^3,
\nonumber \\
 |\Delta \bm{Q}^{(1)}-\Delta \bm{Q}^{(4)}|
&\le& \epsilon^3,
\label{ABF}
\end{eqnarray}
are obtained.
Therefore we obtain
\begin{eqnarray}
 |\Delta \bm{P}^{(1)}_0|
&\le&
 |\Delta \bm{P}^{(4)}_0|
 +  |\Delta \bm{P}^{(1)}_0 - \Delta \bm{P}^{(4)}_0|
\le \epsilon^3, \\
 |\Delta \bm{Z}^{(1)}|
&\le&
 |\Delta \bm{Z}^{(4)}|
 +  |\Delta \bm{Z}^{(1)} - \Delta \bm{Z}^{(4)}|
\le \epsilon^2, \\
 \Biggl|
  {\Delta \bm{Q}^{(1)}_0 \over t}
 \Biggr|
&\le&
 |{\Delta \bm{Q}^{(4)}_0 \over t}|
 +  |{\Delta \bm{Q}^{(1)}_0 \over t} - 
     {\Delta \bm{Q}^{(4)}_0 \over t}|
\le \epsilon^2,
\end{eqnarray}
We have proved (\ref {AAC}).

{\it Lemma}
\begin{eqnarray}
 |\Delta \delta \bm{P}^{(1)} -
  \Delta \delta \bm{P}^{(m)}|
 &\le&
 {\epsilon \over t}
 (
|\delta \bm{X}^{(m)}| |\Delta \bm{X}^{(m)}|
+ |\Delta \delta \bm{X}^{(m)}|
 ), 
\nonumber \\
 |\Delta \delta \bm{Q}^{(1)} -
  \Delta \delta \bm{Q}^{(m)}|
 &\le&
 {\epsilon \over t}
 (
|\delta \bm{X}^{(m)}| |\Delta \bm{X}^{(m)}|
+ |\Delta \delta \bm{X}^{(m)}|
 ), 
\end{eqnarray}
where 
\begin{equation}
 |\Delta \delta \bm{X}^{(m)}|=
  |\Delta \delta \bm{Q}^{(m)}|
+ |\Delta \delta \bm{P}^{(m)}|.
\end{equation}

{\it Proof of Lemma}

By using (\ref {ABB}), we obtain
\begin{equation}
  |\Delta \bm{Q}^{(m)}|
 +|\Delta \bm{P}^{(m+1)}|
\le |\Delta \bm{X}^{(m+1)}|.
\label{ABG}
\end{equation}
In the same way,
\begin{equation}
  |\delta \bm{Q}^{(m)}|
 +|\delta \bm{P}^{(m+1)}|
\le |\delta \bm{X}^{(m+1)}|
\label{ABH}
\end{equation}
is obtained.
We take the difference between
\begin{equation}
 \delta \bm{p}^{(m)}=\delta \bm{p}^{(m+1)}
 +(\delta \bm{q}^{(m)} \cdot {\partial \over \partial \bm{q}^{(m)}}
  +\delta \bm{p}^{(m+1)} \cdot {\partial \over \partial \bm{p}^{(m+1)}}
  )
 {\partial S_m \over \partial \bm{q}^{(m)}},
\end{equation}
and
\begin{equation}
 \delta \bm{P}^{(m)}=\delta \bm{P}^{(m+1)}
 +(\delta \bm{Q}^{(m)} \cdot 
      {\partial \over \partial \bm{Q}^{(m)}}
  +\delta \bm{P}^{(m+1)} \cdot 
      {\partial \over \partial \bm{P}^{(m+1)}}
  )
 {\partial S_m \over \partial \bm{Q}^{(m)}}.
\end{equation}
By the mean value theorem, we get
\begin{eqnarray}
 &&
 (\delta \bm{q}^{(m)} \cdot {\partial \over \partial \bm{q}^{(m)}}
  +\delta \bm{p}^{(m+1)} \cdot {\partial \over \partial \bm{p}^{(m+1)}}
  )
 {\partial S_m \over \partial \bm{q}^{(m)}}
\nonumber \\
 && -  
  (\delta \bm{Q}^{(m)} \cdot {\partial \over \partial \bm{Q}^{(m)}}
  +\delta \bm{P}^{(m+1)} \cdot {\partial \over \partial \bm{P}^{(m+1)}}
  )
 {\partial S_m \over \partial \bm{Q}^{(m)}}
\nonumber \\
 &=&  
  (\delta \bm{Q}^{(m)} \cdot {\partial \over \partial \bm{Q}^{(m)}}
  +\delta \bm{P}^{(m+1)} \cdot {\partial \over \partial \bm{P}^{(m+1)}}
  )
  (\Delta \bm{Q}^{(m)} \cdot {\partial \over \partial \bm{Q}^{(m)}}
  +\Delta \bm{P}^{(m+1)} \cdot {\partial \over \partial \bm{P}^{(m+1)}}
  )
 {\partial S_m \over \partial \bm{Q}^{(m)}}
\nonumber \\
 && +
 (\Delta \delta \bm{Q}^{(m)} 
    \cdot {\partial \over \partial \bm{q}^{(m)}}
 +\Delta \delta \bm{P}^{(m+1)} 
    \cdot {\partial \over \partial \bm{p}^{(m+1)}}
  )
 {\partial S_m \over \partial \bm{q}^{(m)}},
\end{eqnarray}
where the differentiations in the right hand side
are taken at the appropriate values between $(\bm{q}, \bm{p})$ and
$(\bm{Q}, \bm{P})$.
So we obtain
\begin{eqnarray}
 |\Delta \delta \bm{P}^{(m)}
 -\Delta \delta \bm{P}^{(m+1)} |
 &\le&
 ( |\delta \bm{Q}^{(m)} |
 + |\delta \bm{P}^{(m+1)} |
 )
 ( |\Delta \bm{Q}^{(m)} |
 + |\Delta \bm{P}^{(m+1)} |
 )
 {\epsilon^m \over t^m}
\nonumber \\
 && +
 ( |\Delta \delta \bm{Q}^{(m)} |
 + |\Delta \delta \bm{P}^{(m+1)} |
 )
 {\epsilon^m \over t^m}
\nonumber \\
&\le& 
  {\epsilon^m \over t^m}
 |\delta \bm{X}^{(m+1)}| |\Delta \bm{X}^{(m+1)}|
\nonumber \\
&&
 + {\epsilon^m \over t^m}
 (|\Delta \delta \bm{Q}^{(m)}|
+ |\Delta \delta \bm{P}^{(m+1)}|
 ),
\label{ABI}
\end{eqnarray}
where (\ref {ABG}) (\ref {ABH}) are used.
In the same way, we get
\begin{eqnarray}
 |\Delta \delta \bm{Q}^{(m)}
 -\Delta \delta \bm{Q}^{(m+1)} |
&\le& 
  {\epsilon^m \over t^m}
  |\delta \bm{X}^{(m+1)}| |\Delta \bm{X}^{(m+1)}|
\nonumber \\
&&
 + {\epsilon^m \over t^m}
 (|\Delta \delta \bm{Q}^{(m)}|
+ |\Delta \delta \bm{P}^{(m+1)}|
 ).
\label{ABJ}
\end{eqnarray}
From the above inequality,
\begin{eqnarray}
 |\Delta \delta \bm{Q}^{(m)}|
&\le&
  |\Delta \delta \bm{Q}^{(m)}
 - \Delta \delta \bm{Q}^{(m+1)} |
 +|\Delta \delta \bm{Q}^{(m)}|
\nonumber \\
&\le&
 {\epsilon^m \over t^m}
 |\delta \bm{X}^{(m+1)}| |\Delta \bm{X}^{(m+1)}|
+{\epsilon^m \over t^m}
 (|\Delta \delta \bm{Q}^{(m)}|+ |\Delta \delta \bm{P}^{(m+1)}|)
\nonumber \\
&&
+ |\Delta \delta \bm{Q}^{(m+1)}|,
\end{eqnarray}
which yields
\begin{equation}
 |\Delta \delta \bm{Q}^{(m)}| \le 
{\epsilon^m \over t^m}
 |\delta \bm{X}^{(m+1)}| |\Delta \bm{X}^{(m+1)}|
+ |\Delta \delta \bm{Q}^{(m+1)}|
+ {\epsilon^m \over t^m}
  |\Delta \delta \bm{P}^{(m+1)}|.
\end{equation}
Then we get
\begin{equation}
 |\Delta \delta \bm{Q}^{(m)}|+ |\Delta \delta \bm{P}^{(m+1)}|
\le {\epsilon^m \over t^m}
 |\delta \bm{X}^{(m+1)}| |\Delta \bm{X}^{(m+1)}|
 + |\Delta \delta \bm{X}^{(m+1)}|.
\label{ABK}
\end{equation}
By using (\ref {ABK}) to (\ref {ABI}) (\ref {ABJ}),
we obtain
\begin{eqnarray}
  |\Delta \delta \bm{P}^{(m)}
 - \Delta \delta \bm{P}^{(m+1)} |
&\le&
 {\epsilon^m \over t^m}
 (|\delta \bm{X}^{(m+1)}| |\Delta \bm{X}^{(m+1)}|
 + |\Delta \delta \bm{X}^{(m+1)}|)
\nonumber \\
  |\Delta \delta \bm{Q}^{(m)}
 - \Delta \delta \bm{Q}^{(m+1)} |
&\le&
 {\epsilon^m \over t^m}
 (|\delta \bm{X}^{(m+1)}| |\Delta \bm{X}^{(m+1)}|
 +|\Delta \delta \bm{X}^{(m+1)}|)
\end{eqnarray}
By using the above inequalities,
\begin{eqnarray}
 |\Delta \delta \bm{X}^{(m)}|
&\le&
 |\Delta \delta \bm{X}^{(m+1)}|
\nonumber \\
&&
 + {\epsilon^m \over t^m}
 (|\delta \bm{X}^{(m+1)}| |\Delta \bm{X}^{(m+1)}|
 +|\Delta \delta \bm{X}^{(m+1)}|),
\end{eqnarray}
is obtained.
Then we get
\begin{eqnarray}
 |\Delta \delta \bm{P}^{(1)}- \Delta \delta \bm{P}^{(m)}|
&\le&
 |\Delta \delta \bm{P}^{(1)}- \Delta \delta \bm{P}^{(2)}|
+ \cdot \cdot \cdot 
+|\Delta \delta \bm{P}^{(m-1)}- \Delta \delta \bm{P}^{(m)}|
\nonumber \\
&\le&
 {\epsilon \over t}
 (|\delta \bm{X}^{(m)}| |\Delta \bm{X}^{(m)}|
 +|\Delta \delta \bm{X}^{(m)}|).
\end{eqnarray}
In the same way, we obtain
\begin{equation}
  |\Delta \delta \bm{Q}^{(1)}- \Delta \delta \bm{Q}^{(m)}|
\le
 {\epsilon \over t}
 (|\delta \bm{X}^{(m)}| |\Delta \bm{X}^{(m)}|
 +|\Delta \delta \bm{X}^{(m)}|).
\end{equation}
These complete the proof.

{\it Proof End of Lemma}

From the inequalities (\ref {AAA}), we obtain
\begin{equation}
 \Biggl| {\Delta \bm{X}^{(4)} \over t} \Biggr| 
 \le \epsilon^2.
\end{equation}
From the inequalities (\ref {AAU}), we obtain
\begin{equation}
 |\delta \bm{X}^{(4)}| \le |\delta \bm{Q}_0 (1)| t
 + {1 \over \epsilon} \| \delta \bm{P}_0 \| t
 + \| \delta \bm{Z} \| t.
\end{equation}
From the inequalities (\ref {AAB}), we obtain
\begin{equation}
 {1 \over t}
 |\Delta \delta \bm{X}^{(4)}| \le 
  \epsilon^2 |\delta \bm{Q}_0^{(4)} (1)| 
 +\epsilon \| \delta \bm{P}_0^{(4)} \| 
 +\epsilon \| \delta \bm{Z}^{(4)} \| .
\end{equation}
By using the above three inequlities,
we obtain
\begin{eqnarray}
 |\Delta \delta \bm{P}^{(1)}- \Delta \delta \bm{P}^{(4)}|
&\le&
  \epsilon^2 |\delta \bm{Q}_0^{(4)} (1)| 
 +\epsilon \| \delta \bm{P}_0^{(4)} \| 
 +\epsilon^2 \| \delta \bm{Z}^{(4)} \|
\label{ABL} \\ 
 |\Delta \delta \bm{Q}^{(1)}- \Delta \delta \bm{Q}^{(4)}|
&\le&
  \epsilon^2 |\delta \bm{Q}_0^{(4)} (1)| 
 +\epsilon \| \delta \bm{P}_0^{(4)} \| 
 +\epsilon^2 \| \delta \bm{Z}^{(4)} \|,
\label{ABM}
\end{eqnarray}
for $1 \le t \le 1 / \epsilon$.
By combining (\ref {AAB}) (\ref {ABL}) (\ref {ABM}),
we obtain
\begin{eqnarray}
  \Biggl\| {\Delta \delta \bm{Q}_0^{(1)} \over t} \Biggr\| 
  &\le& \epsilon^2 |\delta \bm{Q}_0^{(4)} (1)|
      + \epsilon \| \delta \bm{P}_0^{(4)} \|
      + \epsilon \| \delta \bm{Z}^{(4)} \|,
\nonumber \\
 \| \Delta \delta \bm{P}_0^{(1)} \| 
  &\le& \epsilon^2 |\delta \bm{Q}_0^{(4)} (1)|
      + \epsilon \|\delta \bm{P}_0^{(4)} \|
      + \epsilon^2 \|\delta \bm{Z}^{(4)} \|,
\nonumber \\
 \| \Delta \delta \bm{Z}^{(1)} \| 
  &\le& \epsilon^2 |\delta \bm{Q}_0^{(4)} (1)|
      + \epsilon \| \delta \bm{P}_0^{(4)} \|
      + \epsilon \| \delta \bm{Z}^{(4)} \|.
\end{eqnarray}
This completes the proof of the propositionB.

\section{Appendix; Proof of Proposition $5.3$}

The proof is essentially the same as that of the preceding
proposition 4.1.
So in this proof we estimate the generating function of the 
canonical transformation, and the residual parts which cannot
be eliminated by the canonical transformation only roughly.
Later we omit $m$.
We consider the canonical transformation induced by 
the generating function given by
\begin{equation}
 T_l (\bm{q}^{(l)}_1, \bm{p}^{(l+1)}, t) =
 \sum_{\bm{k}_1 \ne \bm{0} } T_{\bm{k}_1} (\bm{p}^{(l+1)}, t) 
 e^{i \bm{k}_1 \cdot \bm{q}^{(l)}_1},
\end{equation}
where $T_l$ does not depend on $\bm{q}^{(l)}_0$.

The transformed Hamiltonian is 
\begin{eqnarray}
 &&
 H^{(l+1)} = H^{(l)} + {\partial T_l \over \partial t}
\nonumber \\
 &&
\quad \quad
  = {2 \over \epsilon^2} (\bm{\omega} \cdot \bm{p}^{(l+1)})^{1 / 2}
    + {1 \over  \epsilon^2} 
      {1 \over (\bm{\omega} \cdot \bm{p}^{(l+1)})^{1 / 2}} 
       \bm{\omega}_1 \cdot 
            {\partial T_l (\bm{q}_1^{(l)}, \bm{p}^{(l+1)}, t) 
             \over \partial \bm{q}_1^{(l)} }
     + R_1 
\nonumber \\
 &&
\quad \quad
 + D_l (\bm{p}^{(l+1)}, t)
 + R_2 
 + E_l (\bm{q}_1^{(l)}, \bm{p}^{(l+1)}, t)
 + R_3 
\nonumber \\
 && 
\quad \quad
 + B_l (\bm{q}^{(l+1)}, \bm{p}^{(l+1)}, t)
 + R_4 
 + {\partial T_l (\bm{q}_1^{(l)}, \bm{p}^{(l+1)}, t)
   \over \partial t }
 \end{eqnarray}
where $R_1$, $\cdot \cdot \cdot$ $R_4$ are defined as 
\begin{eqnarray}
 &&
 R_1 = {2 \over \epsilon^2} (\bm{\omega} \cdot \bm{p}^{(l)})^{1 / 2}
     - {2 \over \epsilon^2} (\bm{\omega} \cdot \bm{p}^{(l+1)})^{1 / 2}
       - {1 \over \epsilon^2} 
         {\bm{\omega}_1 \over (\bm{\omega} \cdot \bm{p}^{(l+1)})^{1 / 2}} 
         \cdot (\bm{p}_1^{(l)}-\bm{p}_1^{(l+1)}),
\\
 &&
 R_2 = D_l (\bm{p}^{(l)}, t) 
     - D_l (\bm{p}^{(l+1)}, t),
\\
 &&
 R_3 = E_l (\bm{q}_1^{(l)}, \bm{p}^{(l)}, t) 
     - E_l (\bm{q}_1^{(l)}, \bm{p}^{(l+1)}, t),
\\
 &&
 R_4 = B_l (\bm{q}^{(l)}, \bm{p}^{(l)}, t) 
     - B_l (\bm{q}^{(l+1)}, \bm{p}^{(l+1)}, t).
\end{eqnarray}
We determine the generating function $T_l$ so that 
the leading term depending on the slow angle variables
$\bm{q}^{(l)}_1$ can be eliminated;
\begin{equation}
    {1 \over \epsilon^2} 
    {1 \over (\bm{\omega} \cdot \bm{p}^{(l+1)})^{1 / 2}} 
    \bm{\omega}_1 \cdot {\partial T_l (\bm{q}_1^{(l)}, \bm{p}^{(l+1)}, t) 
               \over \partial \bm{q}^{(l)}_1 } =
    - E_l (\bm{q}_1^{(l)}, \bm{p}^{(l+1)}, t).
\end{equation}
When $E_l$ is decomposed as
\begin{equation}
 E_l = \sum_{\bm{k}_1 \neq \bm{0}} e_{\bm{k}_1} (\bm{p}^{(l+1)}, t) 
       e^{i \bm{k}_1 \cdot \bm{q}^{(l)}_1}
\end{equation}
$T_l$ is written as 
\begin{equation}
    T_l =
    \sum_{\bm{k}_1 \neq \bm{0}} i \epsilon^2 
       (\bm{\omega} \cdot \bm{p}^{(l+1)})^{1/2}
       {1 \over (\bm{\omega}_1 \cdot \bm{k}_1)}
       e_{\bm{k}_1} e^{i \bm{k}_1 \cdot \bm{q}^{(l)}_1}.
\end{equation}
Since $T_l$ is roughly estimated as
\begin{equation}
 T_l \sim {\epsilon^l \over t^l},
\end{equation}
the differences between the original variables
and the transformed variables are estimated as
\begin{eqnarray}
 \bm{p}^{(l)}_0 - \bm{p}^{(l+1)}_0 &=& \bm{0},\\
 \bm{p}^{(l)}_1 - \bm{p}^{(l+1)}_1 &\sim& 
 {\epsilon^l \over t^l},\\
 \bm{q}^{(l+1)} - \bm{q}^{(l)} &\sim& {\epsilon^l \over t^l}.
\end{eqnarray}
By using these estimates, we evaluate the residual 
parts $R_i$ and $\partial T_l / \partial t$;
\begin{equation}
 |R_1| \le \Biggl[
  - {1 \over 4} {1 \over \epsilon^2}
    {\bm{\omega}^i_1 \bm{\omega}^j_1 
    \over (\bm{\omega} \cdot \bm{p})^{3/2} }
    (\bm{p}^{(l)}_1-\bm{p}^{(l+1)}_1)^i
    (\bm{p}^{(l)}_1-\bm{p}^{(l+1)}_1)^j
           \Biggr]
    \sim {\epsilon^{2 l} \over t^{2 l}},
\end{equation}
where we use $\bm{\omega}_1 \sim \epsilon$,
\begin{eqnarray}
 |R_2| &\le&
          \Biggl[  
 {\partial D_l (\bm{p}, t) 
           \over \partial \bm{p}_1}
 \cdot (\bm{p}^{(l)}_1 -\bm{p}^{(l+1)}_1) 
          \Biggr]
 \sim {\epsilon^{l+1} \over t^{l+2}},\\
 |R_3| &\le&
          \Biggl[  
 {\partial E_l \over \partial \bm{p}_1}
 \cdot (\bm{p}^{(l)}_1 -\bm{p}^{(l+1)}_1) 
          \Biggr]
 \sim {\epsilon^{2 l-1} \over t^{2 l}},\\
 |R_4| &\le&
          \Biggl[  
 {\partial B_l \over \partial \bm{q}}
 \cdot (\bm{q}^{(l)} -\bm{q}^{(l+1)}) 
 + {\partial B_l \over \partial \bm{p}_1}
 \cdot (\bm{p}^{(l)}_1 -\bm{p}^{(l+1)}_1) 
          \Biggr]
 \sim {\epsilon^{2 (m-1)} \over t^{m}}
      {\epsilon^{l} \over t^{l}},\\
{\partial T_l \over \partial t}
      &\sim& {\epsilon^l \over t^{l+1}}.
\end{eqnarray}
Then $\bm{q}^{(l)}_0$ independent (so $\bm{q}^{(l+1)}_0$ independent)
part is estimated as
\begin{equation}
 R_1 +R_2 +R_3 + {\partial T_l \over \partial t}
 \sim {\epsilon^l \over t^{l+1} },
\end{equation}
while $\bm{q}^{(l)}_0$ dependent (so $\bm{q}^{(l+1)}_0$ dependent)
part is estimated as
\begin{equation}
 R_4 \sim {\epsilon^{2 (m-1)} \over t^{m}}
      {\epsilon^{l} \over t^{l}}.
\end{equation}
The residual part 
\begin{equation}
  R= R_1 +R_2 +R_3 +R_4+{\partial T_l \over \partial t}
\end{equation}
is decomposed into non-oscillating part $R_0$, slowly
oscillating part $R_s$ and fast oscillating part $R_f$;
\begin{equation}
 R=R_0 +R_s +R_f,
\end{equation}
where 
\begin{eqnarray}
 R_0 &=& R_{\bm{0}} (\bm{p}^{(l+1)}, t) 
 \sim {\epsilon^l \over t^{l+1}},\\
 R_s &=& \sum_{\bm{k}_1 \neq \bm{0} } 
   R_{\bm{k}_1}(\bm{p}^{(l+1)}, t) 
   \exp{[i \bm{k}_1 \cdot \bm{q}_1^{(l+1)} ]}
 \sim {\epsilon^l \over t^{l+1}},\\
 R_f &=& \sum_{\bm{k}_0 \neq \bm{0}} 
   R_{\bm{k}_0 \bm{k}_1}(\bm{p}^{(l+1)}, t) 
   \exp{[i \bm{k}_0 \cdot \bm{q}_0^{(l+1)} 
       + i \bm{k}_1 \cdot \bm{q}_1^{(l+1)}
        ]}
 \sim {\epsilon^{2 (m-1)} \over t^{m}}
      {\epsilon^{l} \over t^{l}}.
\end{eqnarray}
We define
\begin{eqnarray}
 D_{l+1} (\bm{p}^{(l+1)}, t) &=& D_l (\bm{p}^{(l+1)}, t)+ R_0,\\
 E_{l+1} (\bm{q}^{(l+1)}_1, \bm{p}^{(l+1)}, t) &=& R_s,\\
 B_{l+1} (\bm{q}^{(l+1)}, \bm{p}^{(l+1)}, t) &=& 
 B_l (\bm{q}^{(l+1)}, \bm{p}^{(l+1)}, t) +R_f,
\end{eqnarray}
By transforming canonically, we succeeded in lowering 
the residual part $E_l \sim \epsilon^{l-1} /t^l$ to
$E_{l+1} \sim \epsilon^l / t^{l+1}$.

\section{Appendix; Proof of Propositions $5.5$, $5.6$}

\subsection{Proof of Proposition $5.5$}

{\bf Lemma}

\begin{eqnarray}
 \Biggl\| {\delta \bm{q}_0 \over t} \Biggr\| &\le&
 |\delta \bm{q}_0 (1)|+ \epsilon |\delta \bm{q}_1 (1)|
 + {1 \over \epsilon^2} |\delta \bm{p}_0 (1)|
 + {1 \over \epsilon} |\delta \bm{p}_1 (1)|,\\
 \| \delta \bm{p}_0 \| &\le&
 \epsilon^4 |\delta \bm{q}_0 (1)|+ \epsilon^4 |\delta \bm{q}_1 (1)|
 + |\delta \bm{p}_0 (1)|
 + \epsilon^3 |\delta \bm{p}_1 (1)|,\\
 \| {\delta \bm{q}_1 \over t} \| &\le&
 \epsilon^3 |\delta \bm{q}_0 (1)|+ |\delta \bm{q}_1 (1)|
 + {1 \over \epsilon} |\delta \bm{p}_0 (1)|
 + |\delta \bm{p}_1 (1)|,\\
 \| \delta \bm{p}_1 \| &\le&
 \epsilon^4 |\delta \bm{q}_0 (1)|+ \epsilon^2 |\delta \bm{q}_1 (1)|
 + \epsilon |\delta \bm{p}_0 (1)|
 + |\delta \bm{p}_1 (1)|.
\end{eqnarray}

{\it Proof}

From the variational equations of $H^{(3,3)}$, 
we obtain 
\begin{eqnarray}
   \Biggl|{\delta \bm{q}_0 \over t} \Biggr| &\le&
 |\delta \bm{q}_0 (1)|
 + {1 \over \epsilon^2} \| \delta \bm{p}_0 \|
 + {1 \over \epsilon} \| \delta \bm{p}_1 \|
 + \epsilon^4 \Biggl\| {\delta \bm{q}_0 \over t} \Biggr\|
 + \epsilon^2 \Biggl\| {\delta \bm{q}_1 \over t} \Biggr\|,\\
 |\delta \bm{p}_0| &\le&
 |\delta \bm{p}_0 (1)|
 + \epsilon^4 \Biggl\| {\delta \bm{q}_0 \over t} \Biggr\|
 + \epsilon^4 \Biggl\| {\delta \bm{p}_0 \over t} \Biggr\|
 + \epsilon^4 \Biggl\| {\delta \bm{q}_1 \over t} \Biggr\|
 + \epsilon^4 \Biggl\| {\delta \bm{p}_1 \over t} \Biggr\|,\\
  \Biggl| {\delta \bm{q}_1 \over t}
  \Biggr| &\le&
  |\delta \bm{q}_1 (1)|
 + {1 \over \epsilon} \| \delta \bm{p}_0 \|
 + \| \delta \bm{p}_1 \|
 + \epsilon^2 \Biggl\| {\delta \bm{q}_1 \over t} \Biggr\|
 + \epsilon^4 \Biggl\| {\delta \bm{q}_0 \over t} \Biggr\|,\\
 |\delta \bm{p}_1| &\le&
 |\delta \bm{p}_1 (1)|
 + \epsilon^2 \Biggl\| {\delta \bm{p}_0 \over t} \Biggr\|
 + \epsilon^2 \Biggl\| {\delta \bm{q}_1 \over t} \Biggr\|
 + \epsilon^2 \Biggl\| {\delta \bm{p}_1 \over t} \Biggr\|
 + \epsilon^4 \Biggl\| {\delta \bm{q}_0 \over t} \Biggr\|.
\end{eqnarray}
From these inequalities, we can deduce the inequalities
of the Lemma by tedious manipulations.

{\it Proof End}

We assume that
\begin{equation}
      |\delta \bm{q}_0 (1)|
 \sim |\delta \bm{p}_0 (1)|
 \sim |\delta \bm{q}_1 (1)|
 \sim |\delta \bm{p}_1 (1)|
 \sim 1,
\end{equation}
since in the linear perturbation, the scale of the 
perturbation variables is arbitrary.
Then we obtain the proposition below.

{\bf Lemma}

\begin{eqnarray}
 \Biggl\| {\delta \bm{q}_0 \over t} \Biggr\| 
 &\le& {1 \over \epsilon^2},\\
 \| \delta \bm{p}_0 \| &\le& 1,\\
 \Biggl\| {\delta \bm{q}_1 \over t} \Biggr\| 
 &\le& {1 \over \epsilon},\\
 \| \delta \bm{p}_1 \| &\le& 1.
\end{eqnarray}
By using the above lemma to the evolution equations
of perturbation variables, we obtain the evaluation of 
proposition $5.5$.

\subsection{Proof of Proposition $5.6$}

Since $S_m \sim \epsilon^{2m} / t^m$, we obtain
\begin{eqnarray}
 |\delta \bm{q}^{(1,1)}- \delta \bm{q}^{(m,1)}| &\le& 
 {\epsilon^2 \over t} |\delta \bm{x}^{(m,1)}|,\\
 |\delta \bm{p}^{(1,1)}- \delta \bm{p}^{(m,1)}| &\le& 
 {\epsilon^2 \over t} |\delta \bm{x}^{(m,1)}|,
\end{eqnarray}
where
\begin{equation}
 |\delta \bm{x}| = |\delta \bm{q}| + |\delta \bm{p}|.
\end{equation}
On the other hand, since $T_l \sim \epsilon^l / t^l$
and $T_l$ does not depend on $\bm{q}_0$, we obtain
\begin{eqnarray}
 |\delta \bm{q}^{(m,1)}-\delta \bm{q}^{(m,l)}| &\le& 
 {\epsilon \over t} |\delta \bm{y}^{(m,l)}|,\\
 |\delta \bm{p}_0^{(m,1)}-\delta \bm{p}_0^{(m,l)}| &=& 0,\\
 |\delta \bm{p}_1^{(m,1)}-\delta \bm{p}_1^{(m,l)}| &\le& 
 {\epsilon \over t} |\delta \bm{y}^{(m,l)}|,
\end{eqnarray}
where 
\begin{equation}
 |\delta \bm{y}| = |\delta \bm{q}_1| + |\delta \bm{p}|.
\end{equation}
By using the above inequalities, we obtain the lemma
below.

{\bf Lemma}

\begin{eqnarray}
 |\delta \bm{q}^{(1,1)}-\delta \bm{q}^{(m,l)}| &\le& 
  {\epsilon \over t} |\delta \bm{y}^{(m,l)}|
 +{\epsilon^2 \over t} |\delta \bm{q}_0^{(m,l)}|,\\
 |\delta \bm{p}_0^{(1,1)}-\delta \bm{p}_0^{(m,l)}| &=& 
 {\epsilon^2 \over t} |\delta \bm{x}^{(m,l)}|  ,\\
 |\delta \bm{p}_1^{(1,1)}-\delta \bm{p}_1^{(m,l)}| &\le& 
  {\epsilon \over t} |\delta \bm{y}^{(m,l)}|
 +{\epsilon^2 \over t} |\delta \bm{q}_0^{(m,l)}|,
\end{eqnarray}

Since 
\begin{eqnarray}
 {1 \over t} |\delta \bm{y}^{(3,3)}| &\le& 
 {1 \over \epsilon},\\
 {1 \over t} |\delta \bm{q}_0^{(3,3)}| 
 &\le& {1 \over \epsilon^2},\\
 {1 \over t} |\delta \bm{x}^{(3,3)}| &\le& 
 {1 \over \epsilon^2},
\end{eqnarray}
we obtain the proposition $5.6$.

\section{Growth Index of Perturbations}

In this appendix, we calculate the growth rates of perturbations
in the first model ($\lambda \phi^2_1 \phi_2$, 
$2 \mu_1 \approx \mu_2$)
and the second model ($\lambda \phi^2_1 \phi^2_2$, 
$\mu_1 \approx \mu_2$)
which are presented in the beginning of \S4,
assuming that $\lambda / \epsilon$ is of order unity.

The fourth order Hamiltonians $H^{(4)}$ of the first model and 
the second model are written by
\begin{eqnarray}
 H &=& \frac{2}{\epsilon} (\bm{\omega} \cdot \bm{P})^{1/2}
      + A + R,\\
 A &=& \frac{\eta}{\epsilon} \frac{1}{t^{\gamma}}
       \frac{1}{(\bm{\omega} \cdot \bm{P})^{1/2}}
       (M +N \cos{k Q_1}),\\
 |R| &\le& \frac{\epsilon}{t^2},       
\end{eqnarray}
where
\begin{eqnarray}
 && \gamma = 1 \qquad k=1 \\
 && M=0 \qquad N=(P_0-2 P_1) P^{1/2}_1,
\end{eqnarray}
and
\begin{eqnarray}
 && \gamma = 2 \qquad k=2 \\
 && M=2 N=(P_0-2 P_1) P_1,
\end{eqnarray}
respectively.
The coefficients of the perturbation equations
\begin{eqnarray}
 \frac{d}{dt}
 \left(
 \begin{array}{c}
  \delta Q_1
 \\
  \delta P_1
 \end{array}
 \right)
&=&
 \left(
 \begin{array}{cc}
  \dfrac{\partial^2 A}{\partial P_1 \partial Q_1}
 & 
  - \dfrac{1}{2}
  \dfrac{\omega^2_1}{\epsilon}
  \dfrac{1}{(\bm{\omega} \cdot \bm{P})^{3/2}}
  + \dfrac{\partial^2 A}{\partial P^2_1}
 \\
  - \dfrac{\partial^2 A}{\partial Q^2_1}
 & 
  - \dfrac{\partial^2 A}{\partial Q_1 \partial P_1}
 \end{array}
 \right)
 \left(
 \begin{array}{c}
  \delta Q_1
 \\
  \delta P_1
 \end{array}
 \right)
\\
 && +
 \left(
 \begin{array}{c}
  -
  \dfrac{1}{2}
  \dfrac{\omega_0 \omega_1}{\epsilon}
  \dfrac{1}{(\bm{\omega} \cdot \bm{P})^{3/2}}
+ \dfrac{\partial^2 A}{\partial P_1 \partial P_0}
 \\
  - \dfrac{\partial^2 A}{\partial Q_1 \partial P_0}
 \end{array}
 \right)
 \delta P_0 + \bm{R},
\\
 \frac{d}{dt} \delta Q_0
 &=&
 \Biggl(
 - \frac{1}{2} \frac{\omega^2_0}{\epsilon}
   \frac{1}{ (\bm{\omega} \cdot \bm{P})^{3/2} }
 + \frac{\partial^2 A}{\partial P^2_0}
 \Biggr)
 \delta P_0
 + \frac{\partial^2 A}{\partial P_0 \partial Q_1}
 \delta Q_1
\\
 &&
 +
 \Biggl(
 - \frac{1}{2} \frac{\omega_0 \omega_1}{\epsilon}
   \frac{1}{ (\bm{\omega} \cdot \bm{P})^{3/2} }
 + \frac{\partial^2 A}{\partial P_0 \partial P_1}
 \Biggr)
 \delta P_1
 + R,
\\
 \delta P_0 &=& \delta P_0 (1),
\\
 |R| &\le& \frac{\epsilon}{t^2}
  (|\delta P_0|+|\delta Q_1|+|\delta P_1|),
\end{eqnarray}
are given by
\begin{equation}
 \Biggl(
 \frac{\eta}{\epsilon} \frac{1}{t^{\gamma}}
 \Biggr)^{-1}
 \frac{\partial^2 A}{\partial P_1 \partial Q_1}
 =
 - k \frac{1}{(\bm{\omega} \cdot \bm{P})^{1/2}}
     \frac{\partial N}{\partial P_1}
     \sin{k Q_1}
 + \frac{1}{2} k
   \frac{\omega_1}{(\bm{\omega} \cdot \bm{P})^{3/2}}
   N \sin{k Q_1}
,
\end{equation}
\begin{eqnarray}
 \Biggl(
 \frac{\eta}{\epsilon} \frac{1}{t^{\gamma}}
 \Biggr)^{-1}
 \frac{\partial^2 A}{\partial P^2_1}
 &=&
   \frac{3}{4}
   \frac{\omega^2_1}{(\bm{\omega} \cdot \bm{P})^{5/2}}
   (M+N \cos{k Q_1})
  - \frac{\omega_1}{(\bm{\omega} \cdot \bm{P})^{3/2}}
    \Biggl(
     \frac{\partial M}{\partial P_1}
    +\frac{\partial N}{\partial P_1} \cos{k Q_1}
    \Biggr)
\nonumber \\
  && +
     \frac{1}{(\bm{\omega} \cdot \bm{P})^{1/2}}
    \Biggl(
     \frac{\partial^2 M}{\partial P^2_1}
    +\frac{\partial^2 N}{\partial P^2_1} \cos{k Q_1}
    \Biggr)
,
\end{eqnarray}
\begin{eqnarray}
 \Biggl(
 \frac{\eta}{\epsilon} \frac{1}{t^{\gamma}}
 \Biggr)^{-1}
 \frac{\partial^2 A}{\partial P_1 \partial P_0}
 &=&
   \frac{3}{4}
   \frac{\omega_0 \omega_1}{(\bm{\omega} \cdot \bm{P})^{5/2}}
   (M+N \cos{k Q_1})
  - \frac{1}{2}
    \frac{\omega_0}{(\bm{\omega} \cdot \bm{P})^{3/2}}
    \Biggl(
     \frac{\partial M}{\partial P_1}
    +\frac{\partial N}{\partial P_1} \cos{k Q_1}
    \Biggr)
\nonumber \\
  && -
     \frac{1}{2}
     \frac{\omega_1}{(\bm{\omega} \cdot \bm{P})^{3/2}}
    \Biggl(
     \frac{\partial M}{\partial P_0}
    +\frac{\partial N}{\partial P_0} \cos{k Q_1}
    \Biggr)
\nonumber \\
  &&
   + \frac{1}{(\bm{\omega} \cdot \bm{P})^{1/2}}
    \Biggl(
     \frac{\partial^2 M}{\partial P_1 \partial P_0}
    +\frac{\partial^2 N}{\partial P_1 \partial P_0} \cos{k Q_1}
    \Biggr),
\end{eqnarray}
\begin{equation}
 \Biggl(
 \frac{\eta}{\epsilon} \frac{1}{t^{\gamma}}
 \Biggr)^{-1}
 \frac{\partial^2 A}{\partial Q^2_1}
 =
 - k^2 \frac{1}{(\bm{\omega} \cdot \bm{P})^{1/2}}
   N \cos{k Q_1}
,
\end{equation}
\begin{equation}
 \Biggl(
 \frac{\eta}{\epsilon} \frac{1}{t^{\gamma}}
 \Biggr)^{-1}
 \frac{\partial^2 A}{\partial Q_1 \partial P_0}
 =
   \frac{1}{2} k 
   \frac{\omega_0}{(\bm{\omega} \cdot \bm{P})^{3/2}}
   N  \sin{k Q_1}
 - k
   \frac{1}{(\bm{\omega} \cdot \bm{P})^{1/2}}
   \frac{\partial N}{\partial P_0} \sin{k Q_1}
,
\end{equation}
and
\begin{eqnarray}
 \Biggl(
 \frac{\eta}{\epsilon} \frac{1}{t^{\gamma}}
 \Biggr)^{-1}
 \frac{\partial^2 A}{\partial P^2_0}
 &=&
   \frac{3}{4}
   \frac{\omega^2_0}{(\bm{\omega} \cdot \bm{P})^{5/2}}
   (M+N \cos{k Q_1})
  - \frac{\omega_0}{(\bm{\omega} \cdot \bm{P})^{3/2}}
    \Biggl(
     \frac{\partial M}{\partial P_0}
    +\frac{\partial N}{\partial P_0} \cos{k Q_1}
    \Biggr)
\nonumber \\
  && +
     \frac{1}{(\bm{\omega} \cdot \bm{P})^{1/2}}
    \Biggl(
     \frac{\partial^2 M}{\partial P^2_0}
    +\frac{\partial^2 N}{\partial P^2_0} \cos{k Q_1}
    \Biggr).  
\end{eqnarray}
Since the correction to the growth rate of perturbations 
by $R$ is of order $\epsilon$ as long as we consider the 
finite small time range, we consider the fixed points with 
dropping $R$.

For simplicity, we consider the case $\omega_1 =0$.

In the first model, around the fixed point 
\begin{equation}
 Q_1 = \frac{\pi}{2} + k \pi \qquad
 2 P_1 = P_0 = c,
\end{equation}
where $k$ is integer, the perturbations are given by
\begin{eqnarray}
 \delta Q_1 &=& 
 t^{(-)^k \Gamma_1} \delta Q_1 (1)
,\\
 \delta P_1 &=& 
 t^{(-)^{k+1} \Gamma_1} 
 (\delta P_1 (1) + \frac{1}{2} \delta P_0 (1))
 - \frac{1}{2} \delta P_0 (1)
,\\
 \delta Q_0 &=&
 \delta Q_0 (1)
 - \frac{1}{2 \epsilon} 
   \frac{\omega^{1/2}_0}{c^{3/2}}
   \delta P_0 (1) (t-1)
\nonumber \\
 &&
 - \frac{1}{2} (t^{(-)^k \Gamma_1}-1) \delta Q_1 (1)
,\\
 \delta P_0 &=& \delta P_0 (1),
\end{eqnarray}
where $\Gamma_1$ is given by
\begin{equation}
 \Gamma_1 = \sqrt{2} \frac{\eta}{\epsilon} 
 \frac{1}{\omega^{1/2}_0}
\end{equation}
The lines $Q_1 = \pi/2 + k \pi$ and $P_1 = c/2$ are the 
heteroclinic orbits.
Since the definition of the Bardeen parameter $\zeta$ contains
the prefactor $1/t$, when $\Gamma_1$ is larger than $1$, 
$\zeta$ grows in proportion to $t^{\Gamma_1 -1}$.
Around the elliptic fixed point
\begin{equation}
 Q_1 = k \pi \qquad 6 P_1 = P_0 = c,
\end{equation} 
the evolution of perturbations is oscillatory.

In the second model, the fixed points are given by
\begin{equation}
 Q_1 = \frac{k}{2} \pi \qquad
 2 P_1 = P_0 = c,
\end{equation}
where for odd $k$ hyperbolic, for even $k$ elliptic.
Around this hyperbolic fixed point the perturbations are 
given by 
\begin{eqnarray}
  \left(
 \begin{array}{c}
 \delta Q_1
 \\
 \delta P_1
 \end{array}
 \right)
 &=&
 c_1 \exp{[\Gamma_2 (1- \frac{1}{t})]}
  \left(
 \begin{array}{c}
 1
 \\
 - \dfrac{\sqrt{2}}{2} c
 \end{array}
 \right)
\nonumber \\
 && +
 c_2 \exp{[-\Gamma_2 (1- \frac{1}{t})]}
  \left(
 \begin{array}{c}
 1
 \\
 \dfrac{\sqrt{2}}{2} c
 \end{array}
 \right)
 + \delta P_0 (1)
  \left(
 \begin{array}{c}
 0
 \\
 \dfrac{1}{2} 
 \end{array}
 \right),
\end{eqnarray}
where
\begin{eqnarray}
 c_1 &=&
  \frac{1}{2} \delta Q_1 (1) 
- \frac{\sqrt{2}}{2} \frac{1}{c} \delta P_1 (1)
+ \frac{\sqrt{2}}{4} \frac{1}{c} \delta P_0 (1)
,\\
 c_2 &=&
  \frac{1}{2} \delta Q_1 (1) 
+ \frac{\sqrt{2}}{2} \frac{1}{c} \delta P_1 (1)
- \frac{\sqrt{2}}{4} \frac{1}{c} \delta P_0 (1),
\end{eqnarray}
and
\begin{eqnarray}
 \delta Q_0 &=&
 \delta Q_0 (1)
 - \frac{1}{2} \frac{1}{\epsilon}
   \frac{\omega^{1/2}_0}{c^{3/2}}
 \delta P_0 (1) (t-1)
\nonumber \\
 &&
 - \frac{5}{32} \frac{\eta}{\epsilon}
   \frac{1}{(\omega_0 c)^{1/2}}
 \delta P_0 (1) (1-\frac{1}{t})
 + R,\\
 \delta P_0 &=& \delta P_0 (1)
\end{eqnarray}
where 
\begin{equation}
 |R| \le \frac{\eta}{\epsilon} \| \delta P_1 \|.
\end{equation}
The growth rate $\Gamma_2$ is given by
\begin{equation}
 \Gamma_2 = \frac{\sqrt{2}}{2} \frac{\eta}{\epsilon}
               \frac{c^{1/2}}{\omega^{1/2}_0}.
\end{equation}
The contribution of the $\eta$ dependent part to the 
Bardeen parameter $\zeta$ 
is proportional to $f(t)$:
\begin{equation}
 f (t) = \frac{1}{t} 
    \exp{[ \Gamma_2 (1- \frac{1}{t}) ]}
\end{equation}
which increases for $t \le \Gamma_2$ and 
\begin{equation}
 \frac{f(\Gamma_2)}{f(1)} =
 \frac{1}{\Gamma_2} \exp{( \Gamma_2 -1 )}.
\end{equation}

In case $\omega_1 \neq 0$, as the time proceeds the term
originating from the unperturbed part becomes dominant 
while the perturbation parts decay as $1 / t^{\gamma}$,
therefore the fixed points disappear.

% \bibliographystyle{/home/usr2/kodama/tex/inputs/jpap}
% \bibliography{cscalar}

\addtolength{\baselineskip}{-3mm}

\end{document}